\documentstyle[11pt,aaspp4]{article}

\begin{document}
 
\title{ON THE DYNAMICAL AND PHYSICAL STATE OF THE `DIFFUSE IONIZED 
MEDIUM' IN NEARBY SPIRAL GALAXIES}
 
\author{Jing Wang$^{1}$ and Timothy M. Heckman$^{1}$}
\affil{Department of Physics and Astronomy, Johns Hopkins University,
Baltimore, MD 21218}

\and 
\author{Matthew D. Lehnert$^{1}$}
\affil{Leiden Observatory, Postbus 9513, 2300RA, Leiden, The Netherlands}

\altaffiltext{1}{Visiting Astronomer, Kitt Peak National Observatory,
     National Optical Astronomy Observatories, which is
     operated by the Association of Universities for Research
     in Astronomy, Inc.\ (AURA) under cooperative agreement
     with the National Science Foundation.}

\begin{abstract}

We report the initial results from a program to study the
morphology, physical state, and kinematics of the `Diffuse Ionized
Medium' (`DIM') in a sample of the nearest and brightest late-type
galaxies. For each of five galaxies (NGC 2403, M 81, NGC 4395, M
51, and M 101) we have analyzed deep narrow-band H$\alpha$ images
of the entire star-forming disk, and long-slit spectra of the inner
($\sim$ 10 kpc) disk with a resolution of 40 to 75 km s$^{-1}$. We
find that the DIM covers most of the star-forming disk, and is
morphologically related to the presence of high-surface-brightness
gas (the giant HII regions). The DIM and the giant HII regions 
differ systematically in their physical and dynamical state. The
DIM is characterized by enhanced emission in the low-ionization
forbidden lines ([OI], [NII], and [SII]), and even the high-ionization
 [OIII]$\lambda$5007 line is moderately strong in the
DIM in at least three cases. This last result contrasts with 
upper limits on the [OIII]
surface brightness in the local DIM of our own Galaxy
(the `Reynolds Layer').
We directly verify the
inference made by Lehnert \& Heckman that the DIM contributes
significantly to the spatially-integrated (global) emission-line
ratios measured in late-type galaxies. We also
find that the DIM is more disturbed kinematically than the gas in
the giant HII regions. The deconvolved (intrinsic) widths of the
H$\alpha$ and [NII]$\lambda$6584 lines range from 30 to 100 km s$^{-
1}$ (FWHM) in the DIM compared to 20 to 50 km s$^{-1}$ in the giant
HII regions. The high-ionization gas in the DIM is more
kinematically disturbed than the low-ionization gas:
the [OIII]$\lambda$5007 lines have
intrinsic widths of 70 to 150 km s$^{-1}$. The differing kinematics
implies that `the DIM' is not a single monolithic phase of the ISM.
Instead, it may consist of a `quiescent DIM' with a low 
ionization-state and small scale-height (few hundred pc) and a `disturbed DIM'
with a high ionization state and moderate scale-height (0.5 to 1
kpc). We argue that the quiescent DIM is most likely photoionized
by radiation from O stars leaking out of giant HII regions
(although this requires fine-tuning the opacity of galactic disks
to ionizing radiation). The disturbed DIM is most likely heated by
the mechanical energy supplied by supernovae and stellar winds.
Since the disturbed DIM accounts for only a minority ($<$ 20\%) of
the H$\alpha$ emission in the regions we have studied, there is no
fundamental energetics problem with this model, but it does
requires mechanically-heated gas to have a high areal covering
factor in the inner disk (which needs to be confirmed observationally).
 We find no clear discontinuity between the
physical and dynamical properties of the giant HII regions and the
quiescent DIM. The quiescent DIM is morphologically related to the
giant HII regions and there is a smooth dependence of the 
emission-line ratios and emission-line widths on the surface brightness of
the emission. Thus, we suggest that a unified approach to the study
of the DIM and giant HII regions in star-forming galaxies will
prove fruitful.

\end{abstract}


\notetoeditor{Fig. 3(a,b) need to be printed in color on glossy stock at the
  end of the issue. Samples of Fig. 3 in color have been sent 
  with our first submission. Fig. 1(a,b,c,d,e) can be printed in 
  black-and-white on text stock within the paper. }

\section{INTRODUCTION}

  The interstellar medium (ISM) in our Galaxy
is known to be a multi-phase
complex system. The most recently-discovered major component of the
ISM is the wide-spread diffuse ionized medium (\cite{rey90,kul88}). 
This gas, usually called the `Reynolds Layer',
is characterized by a relatively low ionization state compared to normal HII
regions, a low surface brightness, relatively large scale height, and 
substantial energetic requirements. These latter are so severe that
the Reynolds Layer must either soak up nearly 100\% of the mechanical energy
supplied by supernovae and stellar winds or the topology of the interstellar 
medium must allow a substantial fraction of the ionizing radiation produced
by massive stars to escape the Galactic disk and propagate to moderate 
scale heights. In either case, the implications for our understanding
of the structure and energetics of the interstellar medium are considerable.

  In several nearby normal spiral galaxies, gas with apparently similar
properties
to the Reynolds Layer
has been found as well (see recent reviews by \cite{det92,ran96conf,walbr96}).
 This gas has been variously referred to
as the `Diffuse Ionized Medium' (DIM), `Diffuse Ionized Gas' (DIG), and
`Warm Ionized Medium' (WIM). In this paper we will adopt the acronym
`DIM'.
In the best studied
case of the edge-on galaxy
NGC 891 (\cite{det90,rankulhes90}), 
a DIM in form of filaments and bubbles
can be detected out to more than a few kiloparsec from the disk mid-plane
(see also Rand 1997). A DIM with similar chaotic structure
 can be seen in some
other face-on late-type galaxies like M 33
and M 31 (e.g. Courtes et al. 1987; Walterbos \& Braun 1994).
This diffuse gas is (like the Reynolds Layer) characterized by the
relative strength of low-ionization emission-lines like 
[OII]$\lambda$3727, [SII]$\lambda\lambda$6717,6731 and [NII]$\lambda$6584
(e.g. Dettmar \& Schulz 1992; Hunter \& Gallagher 1997; Ferguson et al. 1996b). 

More generally,
Lehnert \& Heckman (1994) have shown that the integrated optical spectra of 
a large sample of normal late type galaxies published by Kennicutt (1992a,b)
provide very suggestive evidence that the DIM is generic to such galaxies: 
the global value of the [SII]/H$\alpha$ emission-line
ratio is enhanced by an average factor of about 1.5 to 2.0 relative to
the value characteristic of individual extragalactic HII regions. Based
on the emission-line ratios in the DIM in the Milky Way and NGC 891, they
estimated that the DIM would generically contribute at least 25\% of the
total H$\alpha$ emission in such galaxies.
Indeed, direct H$\alpha$ imaging of several nearby star-forming galaxies has
confirmed this
estimate. The fraction of the DIM
contribution is in the range of
$\sim$20\%--40\%, fairly independent of Hubble type
(ranging from early-type
spiral(Sb) to irregular) and galaxy inclination 
(Kennicutt et al. 1995; 
Ferguson et al. 1996a,b; Hoopes et al. 1996; Hunter \& Gallagher 1997). 

While the DIM is apparently ubiquitous, its dynamical state and its 
ionization and heating
mechanism(s) are still poorly understood.
To improve our understanding of the structure and evolution of
the interstellar medium in our own and other galaxies, we need
answers to questions
such as how the DIM in external galaxies is energized and how much
the DIM in the Milky Way shares in common with the extragalactic
DIMs. Thus, we need  
to verify 
that DIMs are indeed generic in normal
star-forming galaxies, and to elucidate the properties of the DIM 
in such galaxies. We have therefore conducted 
an investigation of a carefully selected sample
of the nearest, largest, and brightest normal late type galaxies. 
We have selected our sample of galaxies according to the following
criteria: 1) Late Hubble type (Sb or later). Such galaxies have relatively
strong and easily studied optical line emission (typical global
H$\alpha$ equivalent widths of 30 to 100$\AA$---cf. Kennicutt 1992a,b).  
The integrated emission-line spectra of such
galaxies contain a substantial contribution from low-ionization gas 
(Lehnert \& Heckman 1994).
2) Nearby (distance $<$ 10 Mpc).
This allows us to obtain images with the highest possible
spatial resolution (1\arcsec\ is typically 20-50 pc), which is
crucial for detecting faint filamentary emission and determining
its morphology and structure. 3) Large angular size (D$_{25} \geq$10\arcmin),
so that we have the maximum number of resolution elements across
the galaxy. 4) Inclination $\leq$ 65$\arcdeg$. We exclude galaxies seen nearly
edge-on where line-of-sight projection effects become severe. 

  In this paper, we report our first results for five galaxies in the sample:
M 51, M 81, M 101, NGC 2403 and 
NGC 4395. Some properties of these galaxies are summarized in Table 1.
Our data consist of both 
narrow-band H$\alpha$ images and long-slit spectroscopic 
observations. The spectra provide information on emission lines from 
H$\alpha$, [SII]$\lambda\lambda$6717,6731, and [NII]$\lambda$6584
in all five galaxies and on
the H$\beta$, [OIII]$\lambda$5007, and [OI]$\lambda$6300 lines in subsets
of the sample.
These data allow us to address the morphology and
global energetics of the DIM, as well as its physical and dynamical 
state. Specifically, we focus on the 
measurements of the DIM contribution
to the global H$\alpha$ luminosity, the physical state of the DIM (as
probed via the relative strengths of the above emission-lines), the
dynamical state of the DIM, and the inter-relationship between these
physical and dynamical conditions.

\section{OBSERVATIONS AND DATA REDUCTION}

  \subsection{H$\alpha$ Imaging Observations}

  Our imaging data were obtained during February 23-28, 1995, with the 
0.9 meter
telescope at KPNO. We used a Tek 2048$\times$2048 CCD with a 0.68\arcsec\  
pixel size, yielding a field of view of 23.2$\times$23.2 arcmin$^2$.
 The H$\alpha$
filter had a FWHM of 26$\AA$ and was centered on 6562$\AA$. 
Thus, relatively little flux from the [NII]]$\lambda\lambda$6548,6584 
was included in these `on band' images. `Off-band' observations of the
continuum were made with a narrow band filter
having a FWHM of 85$\AA$ centered on 6658$\AA$. Several 1800(600) second
exposures were taken through the H$\alpha$ (continuum) filter
for each object.

  Data reduction was done using the IRAF package following standard procedures.
Images were bias-subtracted first and flat-fielded with a master flat combined
from a set of many dome flats. Sky-subtracted images were then geometrically
rectified to align H$\alpha$ on-band images with the off-band images.

The limit to the detection of diffuse, low-surface-brightness
H$\alpha$ emission in our data is set by the accuracy with which the 
contribution of continuum emission in the on-band image
can be subtracted (since the DIM H$\alpha$ emission has a small equivalent
width). In this regard, the images are inferior to long-slit spectra with
a spectral resolution matched to the intrinsic line-widths of the H$\alpha$
emission-lines (to maximize the contrast of the lines against the continuum).
We have therefore done this continuum subtraction in our images
 by scaling the off-band image such that difference image
agreed with the distribution
of the H$\alpha$ surface brightness in the long-slit spectra (see below) in the
region of overlap.

To make this comparison,
we first sliced the continuum-subtracted H$\alpha$ line images 
in the regions covered by the
slit in the relevant spectral observations. We then reproduced
the spatial profile
of H$\alpha$ surface brightness variation in the same regions and compared
to that shown by the spectra (see \S 2.2 for details of making the 
similar profile for H$\alpha$ line in the spectra). By making small iterative 
adjustments to the scaling factor for the off-band image,
 we were able to get good matches of the profiles
between the
imaging and spectral data. This allowed us to obtain the final, most reliable
continuum-subtracted H$\alpha$ images.

Photometric H$\alpha$ imaging
observations of giant
HII regions in these galaxies (Kennicutt 1988) were used to flux-calibrate
our images. We have also used our long-slit spectra to check for
consistency, and find that the agreement in H$\alpha$ flux in the regions
of overlap is
better than
10\% for NGC 2403, NGC 4395 and M 101 and 30\% for M 51. 
This suggests our spectra of H$\alpha$ were
taken under nearly photometric conditions (see below).
 
  Internal extinction in the galaxies has not been corrected for. 
Since the Galactic HI column density in the directions toward these
galaxies is low,
we did not correct the images for foreground Galactic extinction
extinction (it is listed in Table 1).

  \subsection{Optical Spectroscopic Observations}

  Our spectroscopic observations of these five galaxies were made using the
RC spectrograph on the 4 meter telescope at
Kitt Peak National Observatory during February 17-18, 1996, December 2-3, 1996,
December 4-5, 1996, January 31 - February 1, 1997 and February 3-4, 1997.
The T2KB CCD was used as the detector, 
resulting in a scale of 0.69\arcsec\ per pixel. 
A 5\arcmin-long, 2\arcsec-wide slit was placed sampling the nuclei 
as well as representative spiral arm and inter-arm regions of
 our sample galaxies.
 The position angle of the slit for each galaxy is listed
 in Table 2 and shown in Figure 1. Based on HI velocity maps, 
the slit orientation was chosen to be near the kinematic minor axes of the
each galaxy. This was done so that
the velocity gradient of the emission-line gas along the slit should be
minimized (allowing us to sum up the emission from large regions along
the slit without introducing line-broadening due to large-scale velocity
shear). We also insured that the amount of
atmospheric differential refraction across the slit was insignificant
for the particular slit position angles and hour angles of the observations.

Table 2 lists the spectrograph setup and relevant observation information.
One set of data (hereafter the `red spectra') covers the 
wavelength range 
from 6050 to 7090$\AA$ centered at 
H$\alpha$. Because of vignetting in the spectrograph camera and
focus variations across the detector, the useful spectral range is
limited to the region from 6180 to 6950$\AA$.
The data were obtained with the grating KPC-24 in
second order. The resolution as determined from FWHMs of
night sky lines is about 0.8$\AA$ (37 km/s) FWHM at H$\alpha$, 
allowing determination of deconvolved line widths as small as 
$\simeq$20 km/s in data with high signal-to-noise (7:1 or better). This
can be compared with the minimum H$\alpha$ FWHM due to pure thermal
broadening of 22 km/s for T = 10$^{4}$ K.

The second set of data (hereafter the `green spectra') covers the
useful wavelength range from about 4400 to 5300$\AA$, and includes the
H$\beta$ and [OIII]$\lambda$4959,5007 lines. The spectral resolution
as determined from FWHMs of strong emission lines in arc lamp spectra
ranges from 1.0 to 1.5$\AA$ FWHM (60 to 90 km/s). These data
were obtained with either grating KPC-24 or grating KPC-18C in second order.
The green spectrum of N 2403 was taken twice on different nights 
and the data were combined after proper reduction
to achieve high S/N ratio.

The data were reduced following standard procedures of biasing and
flat-fielding. Wavelength calibration was done with the package
TWODSPEC.LONGSLIT
in IRAF, using Thorium-Argon or Helium-Neon-Argon arc lamp exposures.
The uncertainty in wavelength
after calibration for our high resolution spectra
was found less than 0.2$\AA$ by comparing 
wavelengths of night sky lines with standard values.
 
Since the galaxies have 
angular sizes that exceed the slit length, night sky spectra were taken
immediately before and after object exposures, and were used to subtract
the sky. To avoid introducing significant noise, we fit the night
sky spectra in the spatial direction with a low order (no higher than 4th)
polynomial and used
the fits (rather than the actual sky spectra)
to do the sky subtraction.
The relative strengths of night sky lines varied during
the observations in an unpredictable way. Thus, a straightforward method
of subtracting a sky frame from a object frame may
not remove some night sky lines completely, especially near the H$\alpha$,[NII],
and [OI]$\lambda$6300 lines in the red spectra. We instead 
adjusted the scaling factor for sky subtraction interactively in
different wavelength ranges until a visually acceptable subtraction was done.
As a result, different scaling factors were used as necessary to subtract
the sky in the red spectra near the
H$\alpha$, [NII], [SII], and [OI] lines respectively. 
In the green
spectra, the problem was lessened as there are no strong night sky lines 
near H$\beta$ and [OIII]$\lambda$5007. Only small residuals 
were left after sky subtraction and they could be treated
as additional background noise.

Spectra of the standard stars BD +262606, Feige 34, G191B2B and HZ 44 
were taken to do spectrophotometric calibration. For the red
spectra, the comparison between the spectra and our flux-calibrated H$\alpha$
image (see \S 2.1) suggest that the absolute fluxes of these spectra were 
very close to photometric values.
We extrapolated the green spectra to the spectral region covered by our 
red spectra and found both set of spectra show consistent flux scales. 
This indicates that the green data were obtained under photometric
conditions as well.
Cosmic rays near important
emission lines have been removed with IRAF task IMEDIT before
measurements.

We have then constructed the spatial 
profiles
of the H$\alpha$ surface brightness variation along the slit. To do so,
the BACKGROUND task in IRAF was employed 
to fit (in the wavelength direction) 
the stellar continuum in the spectral region near 
H$\alpha$.
To correct for the effect
of the broad underlying stellar H$\alpha$ absorption-line, we used a
4th order polynomial
to model the absorption line profile.
Because the H$\alpha$ emission line is almost 
always very narrow compared to the absorption 
feature, the emission line could be automatically rejected by the 
fitting program and therefore only
the absorption line profile was fitted. The fits were subtracted from the
original data to obtain the pure H$\alpha$ emission-line fluxes. 
Although a polynomial is not an accurate 
representation of the real absorption line profile, this method has been very 
efficient and successful in removing most of the underlying absorption.
The spatial profile of H$\alpha$ surface brightness 
 was then derived from the absorption-corrected spectrum.
The profile was used to locate the regions where the faint gas is, and served
as a reference for the continuum subtraction of our imaging data as well
(\S 2.1).

  There has been no unique way to isolate the DIM from the
rest of a galaxy. Here we simply set an upper limit of 
5$\times$10$^{-17}$ erg s$^{-1}$ cm$^{-2}$ arcsec$^{-2}$ 
to H$\alpha$ surface brightness for  the diffuse gas, corresponding
to an emission measure of 25 cm$^{-6}$ pc at an electron
 temperature of 10$^4$ K. 
We used this limit to separate spectra of the diffuse 
gas from that of HII regions and obtained a set of apertures for
spectra of the DIM.
The red and green spectra were spatially registered
using the positions of a few
HII regions along the slit.
Thus, the H$\alpha$ surface brightness criterion can be applied to 
both sets of spectra, ensuring that the measured line emission of
H$\beta$ and
[OIII] (green data) in a given aperture
is from the same spatial region as
the H$\alpha$, [NII] and [SII] (red data).
The spectra were binned spatially as necessary
for adequate detection of DIM emission lines. 
For very weak lines such as [OI] and [OIII], the spectra had to be 
summed over many 
apertures in order to make a measurement. In these cases, we
selectively added apertures together and avoided apertures
associated with very noisy spectra. This way we could achieve the optimal
S/N ratio to detect the weak lines.

The IRAF task SPLOT 
was used to interactively measure integrated fluxes,
 centroids, and widths of the
emission lines, assuming a single Gaussian for all the emission lines.
The measured linewidths were deconvolved using
the instrumental resolution to give true widths. Since the instrumental
resolution varied slowly as a function of wavelength and
position along the slit, we were careful to use the local value of
the instrumental resolution for this deconvolution (as determined from
night sky lines and the arc lamp exposures).

  Since the absorption features of H$\alpha$ and H$\beta$
are usually very broad (FWHM$>$20$\AA$)
 and shallow while the emission lines are usually very narrow (
FWHM $\leq$2$\AA$),
it is possible to correct for the absorption
in a more rigorous manner than fitting the absorption features
 with a polynomial as discussed above. In most cases, the narrow Balmer
emission lines appear on top of the broad valley of the absorption lines.
It suffices to 
integrate the emission-line flux by simply setting the base of the Balmer
emission lines as the continuum level. In rare cases in which 
the absorption features 
are only moderately broader than emission lines
and further correction needs to be done, 
we compared the wings 
of the absorption features with those in typical A-type star spectra
(Jacoby, Hunter \& Christian 1984).
 A good match
allowed us to estimate how much emission is absorbed underlying the
emission lines.
Because of the narrow Balmer line widths of the DIM, there is negligible
flux missed in the
direct measurement of these emission lines with SPLOT.

  For measurements of the H$\alpha$, [NII]$\lambda$6583, and [SII]
$\lambda$ $\lambda$6717,6731 
emission-line fluxes in the DIM, we summed
spatially along the slit such that 
a S/N ratio of at least 10:1 was achieved in the emission-line fluxes. 
The H$\beta$ and [OIII]$\lambda$5007 lines 
are relatively faint, so regions along the slit were summed to yield
a minimum S/N ratio for these measured 
fluxes of 5:1.
 
We also measured the spectra of HII regions
in the off-nuclear areas, defined to have H$\alpha$ surface
brightness above
2$\times$10$^{-16}$ erg s$^{-1}$ cm$^{-2}$ arcsec$^{-2}$
(EM above 100 cm$^{-6}$ pc). 
In this way we have sampled the brightest HII regions in our spectra,
which can serve as a contrast to the diffuse gas.
The M81 spectra do not cover this type of bright HII regions and were
not measured this way.
In addition, individual HII regions which appeared to be relatively
isolated (not blended with other HII regions)
were selected in order to 
study radial variations of the emission-line ratios [NII]$\lambda$6583/H$\alpha$
and [SII]$\lambda\lambda$6717,6731/H$\alpha$ from the HII region out into the surrounding
DIM. To do so, the
spectra were
extracted with 2--4\arcsec\ long apertures 
marching outward from each HII region. The size of the aperture was
chosen so that the S/N ratios of measured 
emission-line fluxes were at least 5:1.

The precise specifications of the spectroscopic apertures used and the
resulting measurements of line ratios, line widths, and surface
brightnesses for the DIM and HII regions are given in Tables 3 and 4
for the red and green spectra respectively.

\section{RESULTS}

   \subsection{H$\alpha$ Images}

  The H$\alpha$ images of our sample galaxies are shown in Figure 1. 
The images have been smoothed with a box having a size of
2$\arcsec\times$2$\arcsec$.
A faint, diffuse gas component can be seen throughout the disks of 
these galaxies. The detection of extremely faint
gas is affected by the uncertainty of continuum subtraction of
the images. But as discussed in \S 3.2, the wide distribution
of the DIM has also been confirmed by our spectra, which have
a higher sensitivity.
The diffuse gas is preferentially along the edges
of spiral arms and around isolated HII regions, as generally the case
for the DIM (e.g. Ferguson et al. 1996a; Hoopes et al. 1996). 
For M 81, the
bright diffuse gas component within a radius of 2--3\arcmin\ from the nucleus
seems to not be associated with any prominent star-forming regions and
is probably different in nature from the DIM seen elsewhere. The
LINER-type nucleus in this galaxy or a population of hot post-main-sequence
low-mass stars may play a role in energizing the gas (Devereux, Jacoby, \&
Ciardullo 1995, 1996).

  To study the DIM contribution to the total H$\alpha$
flux in our sample, we clipped out discrete HII regions at 
a given surface brightness level from the unsmoothed images and
smoothed the images with a 2$\arcsec\times$2$\arcsec$ box.
Because the average surface brightness of the DIM ($\sim$10 pc cm$^{-6}$)
is comparable to the noise of the background of the resulting clipped 
images, we 
determined the flux from these images by integrating
from $-3\sigma$ of the background of the images to the clipping 
surface brightness level. We repeated this
process over a range in the clipping threshold in surface brightness
(emission measure).
 Figure 2 shows the fraction of total H$\alpha$ flux the gas contributes
as a function of emission measure cut-off. 
 Figure 2 serves only as a rough estimate of
the DIM contribution. The measurement
was not made below a surface brightness corresponding to 3$\sigma$ 
above the background of the unsmoothed images.
This minimum clipping surface brightness level corresponds to an 
emission-measure of  90 pc cm$^{-6}$ for 
M 51 and M 101, 40 pc cm$^{-6}$ for M 81 and NGC 2403, and 60 pc cm$^{-6}$
for NGC 4395. The depths of the images do not significantly affect the 
overall result in Figure 2.

 There are four effects that should be noted when
interpreting Figure 2. First, the light from
HII regions, scattered by the optics in the telescope and camera 
and by the dust in the galaxies, have not been corrected. Hoopes et al. (1996)
showed that this correction is minor for their sample galaxies. Furthermore,
the emission-line ratios of the DIM (see \S 3.2) are significantly 
different from that of typical HII regions. No scattering processes 
can account for this difference. Thus, scattered light is unlikely
dominant in the DIM flux and we
do not consider this effect here. Second, a more reliable determination 
of this contribution
should include a correction of the missed diffuse gas flux on top of
clipped-out HII regions. A simplified method of making this correction
(by assuming the diffuse gas in the clipped-out regions
has the mean surface brightness level measured elsewhere (e.g.
 Ferguson et al. 1996a;
Hoopes et al. 1996)
 may increase the DIM fraction significantly, especially for low cut-off levels.
At 50 pc cm$^{-6}$, the correction factor may be up to 2 (Hoopes et al. 1996).
Third, for the purpose of our study, we did not indicate 
the uncertainty in continuum subtraction in Figure 2.  Based on our comparison
of our spectra and images, this uncertainty is minor at the relevant levels
of emission measure,
and will not alter the essential behavior of the growth
curve. Fourth, no extinction correction has been applied. In a
situation in which HII regions suffer more average extinction than the DIM,
the intrinsic DIM fraction should 
drop accordingly.

  The growth curves shown in Figure 2 vary among our sample galaxies. 
M 51, M 101 and NGC 2403 have an abundance of very bright HII regions 
in their disks. In these 
three galaxies the DIM contributes to the global H$\alpha$ luminosity
at a  relatively  low level. NGC 4395 and M 81 have a  smaller number
of bright HII regions and so the DIM becomes more important globally.
We verified that the high DIM fraction in M 81 is not due to 
the prominent diffuse gas component in the inner disk or bulge. 
To do this test, we masked out a circular region within a radius of 2.5\arcmin\ 
from the nucleus and constructed the growth curve again. The result showed
little difference from that in Figure 2. Thus, the high fraction of the DIM 
contribution is purely due to the fact that M 81 is a quiescent, early-type
spiral. 

  We have compared the growth curves of the three galaxies in the 
Sculptor group shown by Hoopes et al.
(1996) with ours. NGC 300 has a growth curve very 
similar to NGC 4395 and M 81 while
NGC 253 and NGC 55 match our sample galaxies M 51, M 101 and NGC 2403 well.
The variation of the growth curves among galaxies is then not primarily due
to inclination. For example, M 101 and NGC 253 have inclination angles of
17$\arcdeg$ and 78$\arcdeg$ respectively, yet both galaxies show similar growth
curves. The contribution of the DIM to the total H$\alpha$
luminosity is probably a function of global 
star formation strength too. Obviously, the galaxies abundant in
luminous giant HII regions tend to have a smaller fractional contribution 
to H$\alpha$ luminosity from the DIM. 

Although we (and others) have defined the DIM in terms of a surface-brightness
criterion, the D in DIM stands for `Diffuse'. Thus, we have also tried to
characterize the relative importance of the DIM in each galaxy using a
measurement that involves only the structure or morphology of the H$\alpha$
images. While some rigorous approach based on a Fourier analysis of
the images could be attempted, we have just characterized the `DIMness'
of the galaxy by taking the ratio of the mean and the r.m.s. of the H$\alpha$
surface-brightness within the surface area of the galaxy delimited by
the 25 B magnitude per square arcsec isophote. These results (and all
the global properties of the H$\alpha$ images we have measured) are
listed in Table 5. Using the ratio of the mean/r.m.s. as a measure 
of DIMness, we find that M 101 stands out from
the rest of the sample.
It is dominated by very bright giant HII regions that occupy only
a small fraction of the disk surface area, and has a correspondingly small
ratio of mean/r.m.s.

Interestingly, M 51 has a pronounced `diffuse' H$\alpha$
emission, based on its high value of mean/r.m.s. This
morphological criterion thus yields a result that contrasts sharply with
the `growth curve' shown in Figure 2. That is, M 51 has a very small
contribution from gas with low surface-brightness (Figure 2),
but its emission is relatively diffuse (Table 5). It remains to be
seen whether the absolute surface-brightness or the morphology is the more
fundamental and physically-meaningful way of defining the DIM.
   
\subsection{Spectroscopic Data}
        \subsubsection{Overview of Properties of the DIM}
  
Because of the better sensitivity of the
spectral data compared to the images, the spectra can provide us with detailed
information about the global distribution of the very faint DIM.
   One of the most interesting results from our spectroscopic data
is that we can see truly diffuse, low-surface-brightness emission-line
gas almost everywhere along the slit. 

This is illustrated in Figure 3, which shows the sections of the long-slit
spectra containing emission lines of interest, and in Figure 4 which
plots the surface brightness distributions of H$\alpha$, [NII], H$\beta$,
and [OIII] along the
slit for galaxies for which we have data of sufficient quality.
H$\alpha$+[NII] can be easily seen almost everywhere
along the slit except for the [NII] emission in the NGC 4395 spectrum. Thus,
we also include in Figure 3 the [SII] emission for NGC 4395 which shows
detectable emission everywhere. The relative weakness of the [NII]
emission throughout NGC 4395 is most likely an abundance effect (as we will
discuss below).

Reliable detections of the [OIII]$\lambda$5007 and H$\beta$ lines
were made in the DIM in all case except for M 81 (where the stellar background
is very bright). The [OI]$\lambda$6300 line was reliably detected only in the
DIM of NGC 2403 (Figure 3).
A noticeable feature in the green spectra of these
galaxies is that [OIII]$\lambda$5007
for the diffuse gas can sometimes be stronger than 
H$\beta$ in the same spatial regions 
(the opposite of the usual case in the HII regions).
We will discuss this in detail below.

As can be seen in Figure 4, 
faint gas can be clearly seen below our arbitrary H$\alpha$
surface brightness limit for the DIM ($<$
5$\times$10$^{-17}$ erg s$^{-1}$ cm$^{-2}$ arcsec$^{-2}$).
We reject the possibility that the diffuse emission is mainly from
light of HII regions scattered into the line of sight because this is
inconsistent with the pronounced differences in the 
emission-line ratios between the DIM and the HII regions (as we will show
 in the following
section). Thus, our spectroscopic data shows that
the DIM is
apparently ubiquitous throughout the inner disks of late type galaxies. 

   To verify the suggestion made by Lehnert and Heckman (1994) that
the DIM significantly contributes to the integrated spectra of normal
star-forming galaxies, we measured the [NII]/H$\alpha$, [SII]/H$\alpha$,
and [OIII]/H$\beta$
ratios from the integrated spectra. In these we excluded the nuclear
contribution from
within the central $\sim$25\arcsec\ aperture 
because the slit length is
only 5\arcmin\ (so our original spectra may be biased towards the nuclear
spectra, which may not be representative of the galaxy disk). We then find
that for each galaxy the [NII]/H$\alpha$ and [SII]/H$\alpha$ 
ratios from the integrated
spectrum are enhanced relative to the corresponding HII region
values in the same region of these galaxies by a factor from 1.1 to 2, 
while the [OIII]/H$\beta$ ratio shows a range of behavior, but no
systematic trend (Table 6). This is quantitatively consistent with the results
of Lehnert \& Heckman (1994) based on the
analysis of
the Kennicutt(1992a,b) sample.

\subsubsection{Line Ratios of DIM}

Figure 5 shows representative spectra of the DIM in our sample galaxies
near the H$\alpha$, [NII], [SII], and [OIII] lines.
As can be seen in Tables 3 and 6, and Figure 5, the
spectra of the DIM in our sample galaxies have relatively strong
low-ionization forbidden lines of [NII]$\lambda$6584 and 
especially [SII]$\lambda\lambda$6716,6731 compared to 
H$\alpha$. This is typical of the DIM in our own and other
previously-studied galaxies. The contrast between the spectroscopic
properties of the DIM and the high-surface-brightness gas in the HII
regions in these galaxies can clearly be seen in Figure 6, where
we have plotted the ratio of the [SII]/H$\alpha$ lines versus the
ratio of the [NII]/H$\alpha$ for the DIM and HII regions. These
line ratios are typically enhanced by about a factor of three
in the DIM compared to the HII regions (see Table 6).

Note that while there is a strong correlation between [NII]/H$\alpha$
and [SII]/H$\alpha$ in our sample, the
[NII]/H$\alpha$ ratios in NGC 2403 and NGC 4395
are systematically low compared
to the other sample galaxies. 
This is most likely due to 
a lower N/S abundance ratio in the two former galaxies.
As shown by the Vila-Costas and
Edmunds (1993), the origin of nitrogen has both a primary component
and a secondary one, implying the nitrogen abundance scales linearly
with metallicity at low metallicity but quadratically
at medium and high metallicity. This means that the nitrogen abundance is more
sensitive to metallicity variations compared to other heavy elements such
as sulfur, and that the N/O and N/S ratios decrease with decreasing metallicity.
Given the correlation between metallicity and absolute magnitude in
late-type galaxies
(e.g. Zaritsky, Kennicutt, \& Huchra 1994; Skillman, Kennicutt \& Hodge 1989),
it is natural that NGC 4395 and NGC 2403 would have the lowest metallicities
in our sample (see Table 1). In fact, measurements of the O/H
and N/O abundances in the bright HII regions in the inner parts of these
 galaxies
(Villa-Costas \& Edmunds 1993) support
this idea: the N/O ratios ranges from roughly twice solar (M 51), to
roughly solar (M 81 and M 101) to about 40\% solar (NGC 2403 and NGC 4395). 

In our initial analysis we have selected the extremes in surface
brightness in the ionized gas (the DIM and the giant HII regions).
To further explore the behavior of the [NII] and [SII] lines, we
then have measured the relative strengths of these lines as a
function of radius from the center of several giant HII regions in
each galaxy. These HII regions were selected to be bright and 
spatially-isolated from other HII regions to allow us to track the transition
from giant HII region into the surrounding DIM. 
The positions of the selected HII regions relative to the nuclei are 
100\arcsec\ NE and 113\arcsec\ SW for M 51, 117\arcsec\ SE and 82\arcsec\ SE
for M 101, 66\arcsec\ SW for NGC 2403 and 37\arcsec\ NE for NGC 4395.
The results are
shown in Figure 7, where we can see that there is generally a
smooth and monotonic increase in the [SII]/H$\alpha$ and [NII]/H$\alpha$
ratio with increasing distance from the core of the
HII region. This result
emphasizes the {\it continuity} in properties
between the high and low surface-brightness gas.
 
This continuity can be seen explicitly in Figure 8, which plots the dependence
of the [SII]/H$\alpha$ ratio as a function of H$\alpha$ surface
brightness. This plot contains all the regions that we have
measured (the HII regions, the DIM, and the transition regions of
intermediate surface-brightness around the isolated HII regions).
The plot shows that the [SII]/H$\alpha$ ratio is strongly and
inversely correlated with surface-brightness over more than two
orders-of-magnitude in surface-brightness, and for all five
galaxies. The [NII]/H$\alpha$ ratio exhibits a similar behavior, but
the galaxy-to-galaxy scatter is larger, presumably due to the
variation in the N/O abundance ratio among the galaxies (as
discussed above).

The [OI]$\lambda$6300 line---which is a tracer of warm neutral gas, 
and a unique
probe of the DIM---is difficult to detect in our spectra.
The [OI] line in M 81 is coincident with the strong Telluric [OI] line, 
while the [OI] line in NGC 4395 is coincident with another sky line at 
$\sim$6306.9$\AA$. So no
useful measurements in these two galaxies 
were possible. In the other galaxies we have summed
all the DIM data into a single spectrum per galaxy. We then
have made  marginal
(5$\sigma$) detections in M 101 and M 51, and a firm detection in NGC 2403. In
these three cases the relative intensity of the [OI] line is
0.1 to 0.2 of H$\alpha$ (Table 6).
 These ratios are about an order-of-magnitude
larger than in the corresponding
 HII regions ([OI]6300/H$\alpha$
$\approx$0.01--0.03).

While the enhanced strengths of the low-ionization lines due to
[OI], [SII], and [NII] are not surprising in the DIM, our data
imply that the relative strength of the high-ionization
[OIII]$\lambda$5007 line is enhanced in the DIM relative to the
giant HII regions in most of our sample galaxies. This can be seen
in Table 6 and Figure 9. This result is quite surprising if
the DIM is simply photoionized by a dilute radiation field due to
the same population of stars that excite the the giant HII regions
(the `leaky HII region model'). In this case, the enhanced 
low-ionization lines in the DIM would be attributed to a lower
ionization parameter U (the ratio of ionizing photons to electrons)
in the DIM compared to that in the HII regions (cf. Domg\"{o}rgen \& Mathis
1994). The simplest version of these models would therefore require
that the [OIII]/H$\beta$ ratio should drop as we move from the
giant HII regions into the DIM. 
Unfortunately, since the [OIII] surface brightness
is too low in our data, we can not directly test this prediction by plotting
the radial variation of 
the [OIII]/H$\beta$ line ratio across individual HII regions
(as we do for other lines in Figure 7).

\subsubsection{Kinematics}
 
While there have been many recent studies of emission-line ratios
in the DIM, these studies have generally had spectral resolution
that was too low to probe the kinematics of the DIM. In
contrast, our spectra are able to resolve the emission-line widths
throughout the DIM and in most of the giant HII regions. This has
led to several surprising and illuminating results.
 
We begin with a discussion of our results for the H$\alpha$ and
[NII]$\lambda$6584 lines, since these are the two brightest lines
in the DIM and since our red spectra had better velocity resolution
than the green data (instrumental FWHM of typically 40 km/s and 75
km/s respectively). From Figure 10 we see that there is an overall
weak trend for the deconvolved (intrinsic) emission-line widths of
these two lines to inversely correlate with the H$\alpha$ surface
brightness. Typical H$\alpha$ and [NII] emission-line widths range
from 30 to 100 km/s FWHM in the DIM, compared to typically 20 to
50 km/s in the giant HII regions (recall that pure thermal Doppler
broadening at T = 10$^{4}$ K produces a FWHM of about 22 km/s for
Hydrogen). We conclude that the low-ionization gas in DIM is less
quiescent than in the HII regions, but the velocity dispersions are
still small compared to theoretical expectations for gas that has
been shock-heated and pushed to large scale-heights out of the disk
(as we discuss in detail in \S 4).
 
Given the strong correlation we find between the [SII]/H$\alpha$
line ratio and H$\alpha$ surface-brightness (Figure 8), it is not
surprising that there is a trend for the [SII]/H$\alpha$ ratio to
be positively correlated with the H$\alpha$ and [NII] emission-line
widths (Figure 11). The trend is the same for the [NII]/H$\alpha$
ratio, but with more scatter (again, presumably reflecting galaxy-
to-galaxy variations in the N/O abundance ratio). This correlation
between the physical state of the gas (line ratios) and the
dynamical state of the gas (line widths) suggests that mechanical
heating may play some role in exciting the low-ionization lines in
the DIM. However, the best correlation of line ratio is with
surface-brightness (compare Figures 8 and 11).
 
We now turn our attention to the [OIII]$\lambda$5007 emission-line
kinematics. This line is very faint in the DIM and we have had to
sum up large regions of DIM to attain a high enough signal-to-noise
to make measurements of the line widths (see Table 4).
Nevertheless, we find that not only can we spectroscopically
resolve the [OIII] line profiles, but that the deconvolved line widths are
significantly larger than the H$\alpha$ and [NII] lines in the same
region of the DIM (Figure 5). This can be quantitatively evaluated 
in Figure 12: typical
deconvolved [OIII] line widths are 70 to 150 km/s in the DIM, or
about a factor of 1.5 to 2 larger than the [NII] line widths. We emphasize
that the line widths plotted in this figure were measured for {\it
exactly} the same spatial region for the [OIII] and [NII] lines.
Thus, the greater widths of the former can not be attributed to a
systematic velocity gradient over the (by necessity) large
extracted region where the measurements were made.
 
This result has several immediate implications. First, `The DIM'
can not be understood as a single physically and dynamically
homogeneous region. Thus, simple single-component models that
attempt to reproduce all the emission-line ratios in the DIM will
give misleading results. Figure 12 implies that there are at least
two components to the DIM. One region (hereafter the `quiescent
DIM') produces most of the H$\alpha$, [NII]$\lambda$6584, and
[SII]$\lambda$6717,6731 line emission. A second region (hereafter
the `disturbed DIM') produces most or all of the high-ionization
[OIII]$\lambda$5007 line emission. The different kinematics of the
high-ionization and low-ionization material imply that the two
components of the DIM will almost certainly have different spatial
distributions in the direction normal to the galactic disk. The
high-ionization gas (disturbed DIM) should have a significantly larger
vertical scale-height, and therefore be relatively more important
in the `extra-planar' DIM seen in some edge-on disk galaxies. We
will quantify this in \S 4.2 below.
 
Even if the gas in the quiescent DIM has a very low ionization
state, and therefore produces only a negligible amount of
[OIII]$\lambda$5007 emission, the gas in the disturbed DIM must
still produce a non-negligible amount of H$\alpha$ emission. That
is, for any plausible model for the heating of this gas, the
H$\alpha$ emission in the disturbed DIM will be at least 30\% as
strong as [OIII]$\lambda$5007 (see \S 4.1 below). We measure the
H$\alpha$ line to be typically about three times brighter than
[OIII] line in the DIM (Table 6). Thus, the disturbed component of
the DIM ought to contribute at least 10\% to the total H$\alpha$
emission in the DIM. We have therefore reanalyzed our data to see
whether the H$\alpha$ emission-line profile shapes in the DIM can
be understood in terms of a strong narrow component (contributed
by the quiescent DIM) and a weak broader component (contributed by
the disturbed DIM). The modest signal-to-noise in our data allows
us only to say that any broad component can contribute no more than
about 20\% to the total H$\alpha$ line flux,
consistent with an expected minimum value of 10\%.
 
\section{DISCUSSION}
\subsection{The Energy Source for the DIM}

\subsubsection{Photoionization}

The results in \S 3.2.3 suggest that there may be more than one
energy source for the DIM. For the quiescent DIM, the `leaky HII
region' model is appealing on a number of grounds. First,
radiation from massive stars is the most abundant ionization source
in these star-forming galaxies (by about an order-of-magnitude).
Second, there is a morphological correspondence between the DIM and
giant HII regions (Hoopes et al. 1996; Ferguson et al. 1996a and
\S 3.1 above). Third, we have
emphasized the {\it continuity} in the physical and dynamical
properties of the gas in the quiescent DIM and the HII regions
(rather than the existence of any well-defined dichotomy). Fourth,
the quiescent state of this component of the DIM argues against
mechanical heating as the dominant ionization source. Finally, even
simple models of photoionization by dilute stellar radiation can
readily match the relative intensities of the low-ionization
species in the quiescent DIM (as we now show).

The ionization state of photoionized gas is primarily determined
by the ionization parameter (U), defined as the ratio of the
density of ionizing photons and electrons in the gas. The low
ionization state of the quiescent DIM (and the implied low value
for U) is a natural consequence of the `leaky HII region model'. Averaged over
the region interior to R$_{25}$, 
the mean H$\alpha$ surface-brightness of the disks of our galaxies
(including both the DIM and the HII regions) is about $3 \times
10^{-17}$ erg cm$^{-2}$ s$^{-1}$ arcsec$^{-2}$. This corresponds to
a disk-averaged flux of ionizing photons of about 10$^{6}$ s$^{-1}$
cm$^{-2}$ ster$^{-1}$. If we assume that a fraction $f_{DIM}$ of
these reach the DIM, then we find U $= 2\pi \times 10^{6}
f_{DIM}/n_{e}c$ as the characteristic value of U.

We can evaluate $n_{e}$ by taking a temperature appropriate for
photoionized gas (T $\sim$ 10$^{4}$ K), and assuming the DIM is in
pressure-balance with the rest of the ISM. In the solar
neighborhood, P/k $\sim 10^{4}$ cm$^{-3}$ K$^{-1}$. The
surface-mass-density of stars in the inner disks of our galaxies is
about an order-of-magnitude higher than the value in the solar
neighborhood, while the surface-mass-density of the interstellar gas
is roughly similar to the local value. Provided that the scale-height
of this gas is not much less than the scale-height of the stars, simple
considerations of hydrostatic equilibrium then imply that the 
mid-plane pressures in the ISM in the inner disks of our galaxies
will be of-order P/k $\sim 10^{5}$ cm$^{-3}$ K$^{-1}$. Since these higher
pressures are not directly demonstrated observationally, we will
use P/k $\sim 10^{4}$ to $10^{5}$ cm$^{-3}$ K$^{-1}$. We then
obtain U $= 4 \times 10^{-4} f_{DIM}$ to $4 \times 10^{-5} f_{DIM}$
as the characteristic value in the mid-plane.
These mid-plane values are appropriate for the quiescent DIM
since (as we will show in \S 4.2 below) it must have a relatively
small scale-height.

Note that the actual value of U in the DIM will be larger than the
above, since the DIM is spatially-correlated with HII regions and
therefore probably sees a photoionizing flux that is higher than
the disk-average. On the other hand, $f_{DIM}$ is of-order 10$^{-
1}$. The implied range in U is then consistent with photoionization
models that show that the key diagnostic line ratios
[NII]/H$\alpha$ and [SII]/H$\alpha$ peak for values of U in the
range $3 \times 10^{-5}$ to $3 \times 10^{-4}$ (e.g. Sokolowski
1993).

This analysis also reveals why the leaky HII region model
would have difficulties explaining the {\it disturbed} DIM. The
lack of obvious broad wings on the H$\alpha$ or [NII] 
emission-lines (\S 3.2) implies that the disturbed DIM must have a
relatively high ionization state. Photoionization models show that
very strong [OIII]$\lambda$5007 emission (e.g. [OIII]/H$\beta$ $\gg$
1) requires logU $> -2$. Using the same arguments as in the above
paragraph, values this large for U would require $n_{e} < 0.02
f_{DIM}$, or P/k $< 4 \times 10^{2} f_{DIM}\ $ cm$^{-3}$ K$^{-1}$. This
pressure is far below plausible ISM pressures. While the
hydrostatic equilibrium condition implies that the gas pressure
will drop with distance out of the disk plane, the required
pressures are so low that the disturbed DIM would have to be
located more than 6 pressure-scale-heights above the inner disk
(inconsistent with our estimates in \S 4.2 below).

Perhaps the biggest challenge for photoionization models is
explaining the pervasiveness of the DIM. Our new 
long-slit data showing H$\alpha$, [NII], and [SII] emission throughout
the inner disk only exacerbates this problem. Here we would like
to emphasize the curious fact that the opacity of the ISM to
ionizing radiation in these galaxies seems to be `fine-tuned'. That
is, we can imagine two extremes for the fate of ionizing radiation
produced by massive stars. First, it may all be absorbed locally
around the complexes of O stars so that a deep H$\alpha$ image would
reveal small intense `islands' of high-surface-brightness
surrounded by very dark `oceans' that cover most of the disk. This
is manifestly not the case (cf. Figures 1 and 2). The other extreme case
is one in which the ISM is optically-thin to ionizing radiation,
allowing it to freely escape into the galactic halo or
intergalactic space. This also seems far from reality (\cite{lei95,gia97}).
 Instead, the ISM seems to be arranged such
that is optically thin enough to allow the gas throughout a
significant fraction of the ISM to be bathed in the diffuse
radiation of the hot stars, yet optically-thick enough to capture
the great majority of the ionizing radiation. Why is this so? Is
this the result of some subtle feedback loop between star-formation
and the structure and physical state of the ISM?

\subsubsection{Mechanical Heating}

While we can not totally rule out photoionization for the disturbed DIM,
we are lead to consider a totally different ionization source. The
disturbed kinematics of this gas could implicate mechanical heating
of the ISM by supernovae and stellar winds. We therefore consider
two alternative forms for such mechanical heating: radiative shocks
and turbulent mixing layers.

Models of low-density radiative shocks (Shull \& McKee 1979; 
Raymond 1997) show that suitably strong
[OIII]$\lambda$5007 emission (e.g. [OIII]/H$\alpha$ $=$ 1 to 3) is
attained for shock speeds $>$ 100 km s$^{-1}$. A strong shock will
accelerate the shocked material to 3/4 of the shock velocity, so
we would expect the velocity dispersion in the DIM to be $>$ 75 km
s$^{-1}$. The typical measured line widths for the
[OIII]$\lambda$5007 line imply a 3-dimensional velocity dispersion
for the disturbed DIM of 60 to 100 km s$^{-1}$, so the required
shock speeds are plausible.

Turbulent mixing layers (Slavin, Shull, \& Begelman 1993) are the
interface between hot gas flowing past cold clouds, and hence have
a temperature that is intermediate between the cloud and hot
phases. The models yield values of [OIII]$\lambda$5007/H$\alpha$ of
1.5 to 3 for much of the parameter space these authors explored.
This specific range encompasses mixing layers with logT $=$ 5.3 and
5.5 when the `cold' clouds have logT $=$ 4, and mixing layers with
logT $=$ 5.5 when the cold clouds have logT $=$ 2. Cooler mixing
layers do not produce significant [OIII] emission. The turbulent
mixing layer models have one advantage over the shock models in
that they do not demand that the mixing layer necessarily have a
large velocity dispersion. Satisfactorily strong [OIII] emission
relative to H$\alpha$ can be produced when the hot gas moves past
the cold cloud at either 25 km s$^{-1}$ or 100 km s$^{-1}$
(consistent with the observed [OIII] line widths in the DIM). High
relative velocities do lead to enhanced emissivity at a given gas
pressure however (see below).

A potential problem with both the shock and turbulent mixing layers
models (especially the latter) is the low predicted
[OIII]$\lambda$5007 surface brightness for gas having the modest
pressures appropriate to a normal disk galaxy ISM (P/k $\sim$
10$^{4}$ K cm$^{-3}$ to $\sim$
10$^{5}$ K cm$^{-3}$ in the inner disk regions probed by our
spectra). For P/k $=$ 10$^{5}$ K cm$^{-3}$, the predicted
[OIII]$\lambda$5007 surface brightnesses from shocks with v = 100
to 150 km s$^{-1}$ are in the range $3 \times 10^{-18}$ to $7
\times 10^{-18}$ erg cm$^{-2}$ s$^{-1}$ arcsec$^{-2}$. This is similar
to typical [OIII]$\lambda$5007 surface-brightnesses in the DIM of
about $5 \times 10^{-18}$ to $1 \times 10^{-17}$ erg cm$^{-2}$ s$^{-1}$
arcsec$^{-2}$. This agreement would then require the areal 
covering-factor of shocked gas within our slit to be unity. However, the
{\it volume filling-factor} of shocked gas in the galaxy ISM would
be much smaller. The shock models of Shull \& McKee (1979) scaled
to P/k $=$ 10$^{5}$ K cm$^{-3}$ imply that the total thickness of the
layer of [OIII]-emitting gas would be of-order 0.1 pc per 
line-of-sight through the DIM (much less than the thickness of the disk ISM).

The surface-brightness agreement is not so good with the models for
turbulent mixing layers. These predict [OIII]$\lambda$5007 surface
brightnesses of 10$^{-18}$ erg cm$^{-2}$ s$^{-1}$ arcsec$^{-2}$ at best, even
for P/k $=$ 10$^{5}$ K cm$^{-3}$. This is an order-of-magnitude smaller
than the typical observed values in the DIM, and would require the
presence of many ($>$ 10) turbulent mixing layers along a typical
line-of-sight through the galaxy disk.

As noted in \S 1, energizing the entire DIM seems problematic using
only the mechanical energy from supernovae and stellar winds, since
it requires tapping essentially {\it all} of this mechanical
energy. However,the energetic requirements for ionizing the
disturbed DIM alone are substantially less. That is, the disturbed
DIM contributes less than 20\% to the total H$\alpha$ luminosity
of the DIM (\S 3.2.2), at least in the regions probed by our
spectra. Thus, mechanical heating can not be ruled out on simple
energetic grounds.

We have seen that the DIM is very pervasive in the inner disks of
our galaxies, so mechanical heating of the DIM would require that
nearly every line-of-sight through the disk intersect regions of
mechanically-heated gas. It is not clear whether the topology or
`porosity' (McKee \& Ostriker 1977) of the ISM in typical disk
galaxies is consistent with this requirement. Observationally, 
there is no direct evidence for a high covering fraction of
diffuse gas mechanically heated to a temperature of 
$10^5$~K or more in typical galactic disks. 
For example, Heiles 
(1990) and Oey \& Clarke (1997) have concluded that only a small fraction
of the surface area of the disks of the Milky Way, M 31, and M 33
is covered by the hot superbubbles created by supernovae and
stellar winds. On the other hand, the available information about soft
X-ray emission from the nearby spiral galaxies
suggests the existence of diffuse, relatively hot gas (T $\ge10^6$~K), 
but little is known about the corresponding areal covering factor. 
Perhaps the best studied case of diffuse, soft X-ray emitting gas in 
our sample galaxies is M 101. Snowden \& Pietsch (1995) have shown that
in the inner disk of M 101, the gas has T$\simeq 10^{5.8}$~K and a 
covering fraction of-order unity. To heat up the gas to this temperature
requires
shock speeds ($\sim 200$ km s$^{-1}$) that are much larger than the typical 
line widths that we have observed in the DIM. 
Certainly, the hot gas seen in the X-ray data may cool and fall back
into the disk thereby contributing to the DIM.

\subsection{The Dynamics and Inferred Vertical Extent of the DIM}

The observed H$\alpha$ and [NII]$\lambda$6584 line widths in the
DIM range from about 30 to 100 km s$^{-1}$ (FWHM), rather similar
to the line widths in the Reynolds Layer in our own Galactic disk
of $\sim$ 30 to 60 km s$^{-1}$ (Reynolds 1985). Since we have
verified that there are no significant velocity shears along the
slit in the regions where we have made these line-widths
measurements, the broadening must be due to small-scale `turbulent'
motions of the gas. The lines are certainly broader than pure
thermal broadening (which is only 22 km s$^{-1}$ FWHM for HI at T
$= 10^{4}$ K). Of course, the [OIII]$\lambda$5007 line-widths are
larger still: typically 70 to 150 km s$^{-1}$ FWHM.

We can use these line-widths to estimate the corresponding 
scale-height of the emitting gas, assuming that the gas clouds have an
isotropic velocity dispersion and move in a gravitational potential
in which they act as test particles. Using optical surface
photometry of the disks of our sample galaxies, and taking
mass-to-light ratio M$_{tot}$/L = 5 solar units for both V and B bands 
(e.g. Binney \& Tremaine 1987) we
estimate that the typical surface mass-density in the inner disks ranges
from 100 to 400 M$_{\odot}$ pc$^{-2}$. We further assume that the
mass that sets the potential has the form 
$\rho$(z) $\alpha$ exp[-z/H] 
where z is the distance out of the mid-plane and H is the 
scale-height for the mass $\rho$. In the disk of the Milky Way, H $\sim$
350 pc (Freeman 1987). Adopting this value, we have then solved for
the scale height h$_{gas}$ for an ensemble of gas clouds moving in
this potential with a velocity dispersion in the z-direction
$\sigma_{z} = 0.43 \times$ (FWHM). The
results are listed in Table 7.

From this we see that typical scale-heights for the quiescent DIM
(based on H$\alpha$ and [NII]$\lambda$6584 line widths) are 300 to 500 pc,
which is significantly less than the values of 600 to 900 pc for
the Reynolds Layer (Reynolds 1993). This is due in part to the much
larger surface mass-densities in the inner regions of our
galaxies and the associated deeper potential well. The
estimated scale-heights for the dominant (quiescent) component of
the DIM are in fact typically comparable to our assumed scale-height for the stellar
disk (350 pc). Thus, we conclude that {\it the DIM in the inner regions of
our galaxies is not in any sense an `extraplanar' phenomenon}. This
is interesting, since the terms `DIM' (or `DIG' or `WIM'!) and
`extraplanar gas' were often used synonymously in early
investigations. In fact, the quiescent state we find for the
dominant component of the DIM may naturally explain why the DIM
seems to be ubiquitous in late-type galaxies, but striking examples
of extraplanar gas are proving to be rare (e.g. \cite{ran96conf,ran96}).

In Table 7 we also list the derived scale-heights for the
disturbed DIM using our measured [OIII]$\lambda$5007 line widths.
These scale-heights are typically 400 to 800 pc, 
or a factor of about 1.5 to 2  greater
than for the quiescent DIM. Thus the disturbed DIM might more
properly be considered `extraplanar'. In this regard it is
intriguing that Rand (1997) found morphological evidence for two
components in the DIM in NGC 891. Since this galaxy is viewed
almost exactly edge-on, Rand fit the vertical dependence of the
mean H$\alpha$ surface-brightness in terms of the sum of two
exponentials. There is a bright component that provides about 84\%
of the total emission and has a scale-height of 500 pc, and a faint
component that provides the other 16\% of the emission and has a
scale height 5 to 6 times larger. The bright and faint components
must have different kinematics to have such different vertical
distributions, and we suggest that they may correspond respectively to
the quiescent and disturbed components of the DIM that we have
identified kinematically. It would be important in this regard to measure
the dependence of the [OIII]$\lambda$5007/H$\beta$ ratio on distance
out of the mid-plane in NGC 891 to see in the more extended DIM component
has a relatively high ionization state (as we find for the disturbed DIM).
The only published [OIII] measurement (Dettmar 1992) is an upper limit
at a distance of only 0.5 kpc out of the disk (in the region dominated
by Rand's inner DIM component, which we would predict to be of low
ionization).
  
\section{Summary}

We have reported on the initial results from a program to study the
morphology, physical state, and kinematics of the `Diffuse Ionized
Medium' (`DIM') in a sample of the nearest and brightest late-type
galaxies. For each of five galaxies (NGC 2403, M 81, NGC 4395, M
51, and M 101) we have analyzed deep narrow-band H$\alpha$ images
covering essentially the entire star-forming disk (a field diameter
of 23.2 arcmin, or 18 to 53 kpc). These images reach limiting
H$\alpha$ surface-brightnesses of about 8--18 $\times 10^{-17}$ erg
cm$^{-2}$ s$^{-1}$ arcsec$^{-2}$ (corresponding to an emission
measure of about 40--90 cm$^{-6}$ pc). We have also analyzed long-slit
spectra covering a single region in the inner disk (5 arcmin, or
4--12 kpc in extent centered on the galactic nucleus). These
spectra cover the H$\beta$ and [OIII]$\lambda$5007 emission-lines
at a velocity resolution of about 75 km s$^{-1}$ (FWHM) and the
[OI]$\lambda$6300, H$\alpha$, [NII]$\lambda$6584, and
[SII]$\lambda\lambda$6717,6731 emission-lines at a velocity
resolution of about 40 km s$^{-1}$. By summing the spectra over
large spatial regions, we have been able to measure emission-lines
down to a limiting surface-brightness (5 $\sigma$) of about $5
\times 10^{-18}$ erg cm$^{-2}$ s$^{-1}$ arcsec$^{-2}$ (an 
emission-measure of 2.5 cm$^{-6}$ pc for H$\alpha$).
 
As in the case of other similar galaxies (e.g. \cite{ferg96,hoop96})
 we find that
diffuse low-surface-brightness gas (a `DIM') covers most of the
star-forming disk, and is morphologically related to the presence
of high-surface-brightness gas (the giant HII regions). Plots of
the cumulative contribution of gas of progressively higher 
surface-brightness to the integrated H$\alpha$ flux are very similar to the
plots for galaxies in the Sculptor group (\cite{hoop96}).
We note that
the D in DIM stands for `Diffuse', and utilize the ratio of the
mean/r.m.s. H$\alpha$ surface-brightness to quantify this. M 51 is
an example of a galaxy with a pronounced DIM as defined by morphology,
but a weak DIM as defined by absolute surface-brightness (it has
lots of diffuse, high-surface-brightness emission). {\it It is not yet
clear whether surface-brightness or morphology is the more 
physically-meaningful way to specify the DIM.}
 
We find that the DIM and the regions of high H$\alpha$ surface-brightness
 (giant HII regions) differ systematically in their
physical and dynamical state. The DIM is characterized by enhanced
emission (relative to H$\alpha$) in the low-ionization forbidden
lines ([OI], [NII], and [SII]). This agrees with results for the
DIM in other galaxies including our own (cf. \cite{det92}
 and references
therein). However, we also find that the high-ionization
[OIII]$\lambda$5007 line is moderately strong
([OIII]$\lambda$5007/H$\beta$ $\sim$ 1) in the DIM. This contrasts
with upper limits on the [OIII] surface brightness in our own
Galaxy and the prototypical edge-on spiral NGC 891 (\cite{rey90,det92}). We
directly verify the inference made by Lehnert \& Heckman (1994)
that the DIM contributes significantly to the spatially-integrated
(global) emission-line spectra of late-type galaxies as published
by Kennicutt (1992a,b).

We also find that the DIM is more disturbed kinematically than the
gas in the giant HII regions. The deconvolved (intrinsic) widths
of the H$\alpha$ and [NII]$\lambda$6584 lines range from 30 to 100
km s$^{-1}$ (FWHM) in the DIM compared to 20 to 50 km s$^{-1}$ in the
giant HII regions. These can be compared to a width of 22 km s$^{-
1}$ for pure thermal broadening of H$\alpha$ at T $=10^{4}$ K. The
high-ionization gas in the DIM is more kinematically disturbed than
the low-ionization gas: in the same regions in the DIM we measure
the [OIII]$\lambda$5007 lines to have typical intrinsic widths of
70 to 150 km s$^{-1}$.
 
The differing DIM kinematics seen in the [OIII] and [NII] lines along
the same line-of-sight
implies that {\it the DIM is not a single
monolithic phase of the ISM}. Instead we propose that there is a
`quiescent DIM' that is responsible for the majority ($>$ 80\%) of
the emission in lines like H$\alpha$, H$\beta$, [NII]$\lambda$6584,
and [SII]$\lambda\lambda$6717,6731, and a `disturbed DIM' that is
responsible for the bulk of the [OIII]$\lambda$5007 emission.
Considerations of hydrostatic equilibrium in the vertical potential
well of the inner disk then imply that the quiescent DIM has a
modest scale-height of several hundred pc (probably similar to the
scale-height of the old stars). Thus, {\it the terms `DIM' and
`extraplanar gas' are by no means synonymous}. The estimated 
scale-heights for the disturbed DIM are about a factor of two larger (0.3
to 1 kpc). This material would have a higher characteristic
ionization state than the quiescent (thin-disk) DIM.
 
We find that the standard `leaky HII region model' in which the DIM
is photoionized by the diffuse Lyman continuum radiation field from O
stars (cf. Domg\"{o}rgen \& Mathis 1994; Sokolowski 1993) can naturally
account for the measured properties of the quiescent DIM. The
ubiquity of faint H$\alpha$ emission (which our spectra show to
fill essentially the entire inner disk) then leads to an apparent need for some
fine-tuning of the physical state of the ISM: {\it the disk must be
optically-thick enough to capture most of the ionizing photons, yet
optically-thin enough to allow these photons to permeate the disk.}
Why is this so? Is there some feedback loop involving massive stars
and the ISM?
 
In contrast, mechanical heating (which is ultimately driven by the
energy supplied by supernovae and stellar winds) is the most
natural ionization source for the disturbed DIM. Either radiative
shocks (with a shock speed $>$ 100 km s$^{-1}$) or turbulent mixing
layers (Slavin, Shull, \& Begelman 1993) can produce relatively
strong [OIII]$\lambda$5007 emission. Since the disturbed DIM
accounts for only a minority ($<$ 20\%) of the H$\alpha$ emission
in the regions we have studied, there is no fundamental energetics
problem in either case. However, the observed surface brightness of
the disturbed DIM requires that essentially every line-of-sight
through the inner disk must encounter at least one shock or ten
turbulent mixing layers. Direct observational evidence for
such a large areal covering factor of mechanically-heated gas in typical
galaxy disks is mixed
(Heiles 1990 and Oey \& Clarke 1997; but see Snowden \& Pietsch 1995). 
 
We also have stressed that {\it there is no clear discontinuity in
physical and dynamical properties of the giant HII regions and the
quiescent DIM}. The quiescent DIM is morphologically related to the
giant HII regions and there is a smooth dependence of the 
emission-line ratios and emission-line widths on the surface brightness of
the emission. In other words, as one moves outward from the HII
regions into the DIM, the properties of the ionized gas change in
a continuous, regular way. The present data on the disturbed DIM
are too limited to make firm statements in this regard, but we are
currently investigating this in a much more extensive new data set.
In any case, we suggest that a unified view of the warm ionized gas in the
disks of star-forming galaxies is likely to be productive.

\acknowledgments
T.M.H. acknowledges partial support of this research by NASA (grant NAGW-3138).
NOAO has partly supported the thesis project of J.W. by covering 
the expenses of his observing trips. 
The work of M.D.L. at Leiden is supported by a program funded by the Dutch
Organization for Research (NWO) and the Dutch Minister of Education and
at IGPP/LLNL (where portions of this work were completed) under the
auspices of the US Department of Energy under contract W-7405-ENG-48.
This research has made use of the NASA/IPAC
Extragalactic Database (NED), which is operated by the 
Jet Propulsion Laboratory, Caltech,
under contract with the National Aeronautics and Space Administration. 
We thank R.Dettmar, A.Ferguson, R.Rand, M.Voit, R.Walterbos and R.Wyse 
for helpful discussions. J.Raymond and J.Slavin have kindly provided 
us some results of their model calculation.
J.W. thanks R.Gonzalez-Delgado for help with the Feb.\ 1996 observing 
run at KPNO.

\clearpage

\begin{figure}
\caption{ Continuum-subtracted H$\alpha$ emission-line images in grayscale. North is to the top and
East is to the left. The slit position we used for each galaxy is shown.
The scale of the images is represented by the length of the slit of 5\arcmin\ .
$a$) M 51, $b$) M 81, $c$) M 101, $d$) NGC 2403, and $e$) NGC 4395. }
\end{figure}

\begin{figure}
\caption{ The growth curve of fractional H$\alpha$ flux
as a function of emission measure cutoff for our five galaxies. 
Symbols are: stars---M 101, triangles---M 51, solid circles---M 81,
crosses---NGC 4395, open circle---NGC 2403. See the text
for details about the construction of the curves. }
\end{figure}

\begin{figure}
\caption{ Grayscale representation of the DIM 
2D spectra. The $y$-axis is the wavelength coordinate
increasing from top to bottom.
The $x$-axis is along the slit. From left to right, the axis 
points towards 
NW for M 101 and towards SW for the rest galaxies. The length of
each window is about 5\arcmin.\ 
a) The strong emission lines [NII]$\lambda$6548, H$\alpha$ and
[NII]$\lambda$6584. The 
[SII]$\lambda\lambda$6717,6731 lines for NGC 4395 are
included to compliment the weak 
[NII] lines. b) The weaker lines H$\beta$,
[OIII]$\lambda$5007 and [OI]$\lambda$6300. The data for those galaxies
that have poor S/N in these emission-lines are not shown. }
\end{figure}

\begin{figure}
\caption{ a) Spatial profiles of the
H$\alpha$ (solid line) and 
[NII]$\lambda$6584 (dotted line) 
surface brightness along the slit. One unit along the $y$-axis
corresponds to 1$\times$10$^{-16}$ erg s$^{-1}$ cm $^{-2}$ arcsec$^{-2}$.
The coordinate of the $x$-axis 
increases from SE to NW for M 101 and from NE to SW for the other galaxies.
The origin of the $x$-axis is set at the nuclear position.
The length of the $x$-axis is about 5\arcmin.\
The upper dashed 
line indicates the cutoff H$\alpha$ surface brightness of 
5$\times$10$^{-17}$ erg s$^{-1}$ cm $^{-2}$ arcsec$^{-2}$ 
that we used to isolate the DIM, and the
lower dashed line is the zero level. b) Same as a) but for the spatial
profile of the
[OIII]$\lambda$5007 surface brightness (thick solid line). The H$\alpha$ 
profiles shown in Figure 4$a$ are arbitrarily scaled and represented by 
thin solid line for comparison. The data for those galaxies that have 
poor S/N near [OIII] are not shown. }
\end{figure}

\begin{figure}
\caption{ a) Representative DIM spectra near H$\alpha$,[NII]$\lambda$6583,
and [SII]$\lambda \lambda$6717,6731.
b) Representative [OIII]$\lambda$5007 line profiles compared with [NII]6583
linewidths (illustrated as horizontal bars)
 that have been convolved to have the same instrumental resolution as
the [OIII] data. The [NII]6583 lines were measured from the same spatial regions
where we measured the [OIII] lines.}
\end{figure}

\begin{figure}
\caption{ $Left\ panel$: The [SII]$\lambda \lambda$6717,6731/H$\alpha$
line ratio vs. the [NII]$\lambda$6584/H$\alpha$ line ratio.
Data are from the DIM, bright HII regions, and intermediate
regions around isolated HII regions as shown in Figure 7. 
The DIM points lie to the upper
right, the HII regions to the lower left, and the intermediate points
lie in between (see $right\ panel$). 
Data points are grouped for each galaxy and do not distinguish between
the DIM and bright HII regions.
Symbols are: stars---M 101, triangles---M 51, solid circles---M 81,
crosses---NGC 4395, open circles---NGC 2403. 
The M 81 spectrum does not cover any bright HII regions so
we only include measurements for the faint HII regions.
$Right\ panel$: Same as $left\ panel$ except that the data points are grouped by H$\alpha$ surface
brightness(SB). Open square---SB below 
5$\times$10$^{-17}$ erg s$^{-1}$ cm $^{-2}$ arcsec$^{-2}$, which is the 
cutoff surface brightness we adopted to isolate the DIM, 
solid square---SB above 
2$\times$10$^{-16}$ erg s$^{-1}$ cm $^{-2}$ arcsec$^{-2}$, which is the 
lower limit we used to select bright HII regions, square with cross in
center---SB below 2$\times$10$^{-16}$ erg s$^{-1}$ cm $^{-2}$ arcsec$^{-2}$
but above 5$\times$10$^{-17}$ erg s$^{-1}$ cm $^{-2}$ arcsec$^{-2}$. }
\end{figure}

\begin{figure}
\caption{ Radial variation of the [SII]$\lambda \lambda$6717,6731/H$\alpha$
 and [NII]$\lambda$6584/H$\alpha$ line ratios across individual HII regions. 
M 81 is not included due to its lack of bright HII regions. The line ratios
are shown to increase smoothly from typical HII region values to DIM values 
as the curves move
from the centers of HII regions to faint, surrounding gas. 
Solid line---M 101; Dotted line---M 51; Dashed line---NGC 2403;
Dot-dashed line---NGC 4395.}
\end{figure}

\begin{figure}
\caption{ The [SII]$\lambda\lambda$6717,6731/H$\alpha$ line ratio vs.
H$\alpha$ surface brightness. 
Data are from the DIM, bright HII regions, and intermediate
regions around isolated HII regions as shown in Figure 7.
Data points are grouped for each galaxy and do not distinguish between the
DIM and bright HII regions.
Symbols used are the same as Fig. 6: 
stars---M 101, triangles---M 51, solid circles---M 81,
crosses---NGC 4395, open circles---NGC 2403. 
The DIM points lie to the upper
left, the HII regions to the lower right, and the intermediate points
lie in between. The M 81 spectrum does not cover any bright HII regions so
we only include measurements for the faint HII regions.}
\end{figure}

\begin{figure}
\caption{ Line ratio diagrams. Solid squares---bright HII region data
from M 51, M 101, NGC 2403 and NGC 4395;
The DIM data points are represented in the same convention adopted 
in the previous figures: stars---M 101, triangles---M 51,
large crosses---NGC 4395, open circles---NGC 2403. Arrows mean
that [OIII]$\lambda$5007 is below our 5$\sigma$ detection limit.
Sokolowski(1993)'s standard photoionization model (solid curve) and
dust depletion model (dotted curve) are also shown. Small crosses
on the curves indicate the value of the ionization parameter U (from
lower-right to upper-left along each curve): 3$\times$10$^{-4}$,
5$\times$10$^{-4}$,
7$\times$10$^{-4}$, 9$\times$10$^{-4}$,
2$\times$10$^{-3}$, 4$\times$10$^{-3}$,
6$\times$10$^{-3}$. The high [OIII]/H$\beta$ point for NGC 4395 is for 
gas near the Seyfert nucleus. }
\end{figure}

\begin{figure}
\caption{ H$\alpha$ surface brightness vs. deconvolved line width.
The line widths are
measured from the H$\alpha$ line if the EQW(H$\alpha$) is larger
than 3$\AA$ and from [NII]$\lambda$6584 if the EQW(H$\alpha$) is less
than 3$\AA$ (to avoid the effect of stellar H$\alpha$ absorption).
The dashed line marks
the lower limit (20 km s$^{-1}$) below which line widths can not be
determined reliably.
Data are for the DIM, bright HII regions, and intermediate
regions around isolated HII regions as shown in Figure 7.
The data point notation is the same as in the previous figures: 
stars---M 101, triangles---M 51, solid circles---M 81,
crosses---NGC 4395, open circles---NGC 2403. The DIM points lie to the lower
right, the HII regions to the upper left, and the intermediate points
lie in between. The M 81 spectrum does not cover any bright HII regions so
we only include measurements for the faint HII regions.}
\end{figure}

\begin{figure}
\caption{The [SII]$\lambda\lambda$6717,6731/H$\alpha$ line ratio vs.
deconvolved line width. The line widths are
measured from H$\alpha$ line if the EQW(H$\alpha$) is larger
than 3$\AA$ and from [NII]$\lambda$6584 if the EQW(H$\alpha$) is less
than 3$\AA$. The dashed line marks
the lower limit (20 km s$^{-1}$) below which line widths can not be
determined reliably.
Data are for the DIM, bright HII regions, and intermediate
regions around isolated HII regions as shown in Figure 7.
The data point notation is the same as in the previous figures: 
stars---M 101, triangles---M 51, solid circles---M 81,
crosses---NGC 4395, open circles---NGC 2403. The DIM points lie to the upper
right, the HII regions to the lower left, and the intermediate points
lie in between. The M 81 spectrum does not cover any bright HII regions so
we only include measurements for the faint HII regions.}
\end{figure}

\begin{figure}
\caption{The deconvolved FWHM of the [OIII]$\lambda$5007 line vs.
the deconvolved FWHM
of the [NII]$\lambda$6584 line for regions in the DIM. 
The solid line is the locus where FWHM([NII]$\lambda$6584) = FWHM(OIII]$\lambda$5007). The data point notation is the same as in the previous figure:
stars---M 101, triangles---M 51,
crosses---NGC 4395, open circles---NGC 2403. 
The arrows imply the [OIII] linewidth can not be determined reliably
 based on our
estimated minimum resolvable linewidth of 40 km s$^{-1}$ for the 
[OIII] line. }
\end{figure}

\clearpage


\begin{planotable}{lcccccc}
\tablewidth{38pc}
\tablecaption{Galaxy Parameters} 
\tablehead{
\colhead{Object }      & \colhead{Hubble Type\tablenotemark{a}} &
\colhead{$i$\tablenotemark{b}} & \colhead{D(Mpc)\tablenotemark{c}}    &
\colhead{$v$\tablenotemark{d}}   &  \colhead{A$_{{\tt H}\alpha}$\tablenotemark{e} } &
\colhead{M$_{B,T}$\tablenotemark{f}  }
}

\startdata

M 51     &  SbcI-II  &       64\arcdeg   &     8.4     &   464   &  0.06  &
-21.0 \nl
M 81     &  SbI-II     &       60\arcdeg   &    3.6     &   -36   &  0.19  &
-20.8  \nl
M 101    &  ScdI     &       17\arcdeg   &     7.4     &   251  &  0.06  &
-21.5 \nl
NGC 2403 &  ScIII     &       62\arcdeg   &     3.2     &   131  &  0.21   &
-19.2 \nl
NGC 4395 &  SdIII-IV   &    38\arcdeg   &     2.6     & 317   &  0.06  & 
-16.7 \nl

\tablenotetext{a}{ From Revised Shapley-Ames Catalog(Sandage \& Tammann 1987).}
\tablenotetext{b}{ Inclination data are from Tully(1988) except the 
        inclination for M 101
        is from Zaritsky et al. (1990). }
\tablenotetext{c}{ Distances are from Feldmeier et al. (1997) (M 51),
        Freedman et al. (1994) (M 81), Kelson et al. (1996) (M 101),
        Karachentsev \& Makarov (1996) (NGC 2403) and Rowan-Robinson (1985)
        (NGC 4395). }
\tablenotetext{d}{ Weighted mean observed heliocentric radial velocity from
        Revised Shapley-Ames Catalog(Sandage \& Tammann 1987).}
\tablenotetext{e}{ Galactic extinction in magnitude at H$\alpha$,
        estimated based on foreground Galactic HI column
        density(Stark et al. 1992) in the direction of the objects. 
        The conversion is done by assuming
        N$_{HI}$/E$_{B-V}$ = 5$\times$10$^{21}$ cm$^{-2}$ mag$^{-1}$ and 
        A$_{{\tt H}\alpha}$ = 2.1 E$_{B-V}$. }
\tablenotetext{f} { Total absolute magnitude in B band from 
        Revised Shapley-Ames Catalog(Sandage \& Tammann 1987), 
        adjusted to our adopted distance.}

\enddata

\end{planotable}



\begin{deluxetable}{lcccccccc}
\tablewidth{0pt}
\footnotesize
\tablecaption{Spectroscopic Observations\tablenotemark{a}}
\tablehead{
\colhead{Object}      &  \colhead{$\lambda_{c}$} & 
\colhead{Coverage\tablenotemark{b}} & \colhead{Grating\tablenotemark{c}} & 
\colhead{Resolution} & 
\colhead{PA} &
\colhead{Blocking Filter} & \colhead{Standard Star} & \colhead{Date} \\
\colhead{} & \colhead{(\AA)} & \colhead{(\AA)} & \colhead{} & 
\colhead{(\AA)} & \colhead{($\arcdeg$)} & \colhead{} & \colhead{} & \colhead{} }

\startdata

 M 51     &  6563 & 6180-6950 & KPC-24  & 0.8-1.0 &  60 & GG495 & BD +262606
& 02/17/96 \nl
        &  4850 & 4450-5250 & KPC-24  & 1.2-1.5 & 60 & BG39 & 
Feige 34 & 12/04/96 \nl
M 81     &  6563 & 6180-6950 & KPC-24  & 0.8-1.0 &  60 & GG495 & BD +262606
& 02/17/96 \nl
        &  4800 & 4360-5230 & KPC-18C & 1.2-1.5 & 60 & GG385+CUSO4 & 
Feige 34 & 12/02/96 \nl
M 101    &  6563 & 6180-6950 & KPC-24 & 0.8-1.0 & 140 & GG495 & BD +262606 & 02/17/96 \nl
        &  4900 & 4500-5300 & KPC-24 & 1.0-1.2 & 140 & 4-96 & 
HZ 44 & 01/31/97 \nl
NGC 2403 &  6563 & 6180-6950 & KPC-24 & 0.8-1.0 &  35 & GG495 & BD +262606
& 02/17/96 \nl
        &  4800 & 4360-5230 & KPC-18C & 1.2-1.5 & 35 & GG385+CUSO4 & 
Feige 34 & 12/02/96 \nl
        &  4900 & 4500-5300 & KPC-24 & 1.0-1.2 & 35 & 4-96 & 
G191B2B & 02/03/97 \nl
NGC 4395 &  6563 & 6180-6950 & KPC-24 & 0.8-1.0 &  60 & GG495 & BD +262606
& 02/17/96 \nl
        &  4800 & 4360-5230 & KPC-18C & 1.2-1.5 & 60 & GG385+CUSO4 & 
Feige 34 & 12/02/96 \nl

\tablenotetext{a}{All observations were made with the KPNO 4m telescope using
the RC spectrograph and the T2KB CCD. 
A slit of 5\arcmin\ long and 2\arcsec\ wide is used 
centered on the nucleus of each of the objects. }
\tablenotetext{b}{Useful spectral range estimated from 1500 pixels of the CCD. }
\tablenotetext{c}{All gratings are set in second order.}
\enddata

\end{deluxetable}



\begin{planotable}{lccccc}
\footnotesize
\tablewidth{40pc}
\tablecaption{Emission-line Properties of DIM Part I. H$\alpha$, [NII] and [SII]} 
\tablehead{
\colhead{Object }     &  \colhead{Aperture}   &  \colhead{$\Sigma_{{\tt H}\alpha}$}
 &  \colhead{$\Delta V_{{\tt H}\alpha/{\tt [NII]}}$}   &  \colhead{[NII]/H$\alpha$} &
\colhead{[SII]/H$\alpha$} \\
\colhead{}  &  \colhead{}  & \colhead{(erg s$^{-1}$ cm$^{-2}$ arcsec$^{-2}$)} &
\colhead{(km s$^{-1}$)}  & \colhead{}  &  \colhead{}
}

\startdata

M 51  &    92.5-90.5\arcsec\ NE  &  3.2$\times$10$^{-17}$  &    91  &    1.7   &   1.1   \nl 
    &  84.9-80.8\arcsec\ NE &  3.5$\times$10$^{-17}$  &     89   &   1.4  &    0.97  \nl
    &   63.5-52.6\arcsec\ NE   & 3.3$\times$10$^{-17}$ &      120    &   1.2  &    0.66  \nl
    &    46.9-62.0\arcsec\ SW &  2.2$\times$10$^{-17}$   &    77   &    1.4   &   0.79 \nl
    &    95.2-102.0\arcsec\ SW &  3.8$\times$10$^{-17}$   &   78   &     1.1  &    0.76  \nl
     &  105.5-107.5\arcsec\ SW &  4.4$\times$10$^{-17}$   &    86   &   0.84   &    0.37  \nl
    &   117.3-152.4\arcsec\ SW  &  1.3$\times$10$^{-17}$   &    83   &    1.5   &   0.87  \nl

\nl

M 81 &  144.9-125.6\arcsec\ NE  &  1.3$\times$10$^{-17}$   &       97   &    1.1   &    1.1  \nl
   &   125.5-119.4\arcsec\ NE &  2.2$\times$10$^{-17}$   &    137  &     1.4  &     1.4  \nl
   &   119.3-93.9\arcsec\ NE  &  8.7$\times$10$^{-18}$    &      63   &    1.3  &     1.6  \nl
   &   93.8-51.1\arcsec\ NE  & 1.8$\times$10$^{-17}$    &   120   &    2.1   &    1.3  \nl
   &   51.0-47.0\arcsec\ NE  & 6.3$\times$10$^{-17}$   &    146   &    2.0   &    1.5  \nl
   &   46.9-40.1\arcsec\ NE &   4.3$\times$10$^{-17}$   &    170  &     2.6  &     1.6  \nl
   &   36.5-29.0\arcsec\ NE &   6.0$\times$10$^{-17}$    &   108   &    1.9  &     1.5  \nl
   &    20.8-28.9\arcsec\ SW &  4.9$\times$10$^{-17}$    &   120   &    2.0  &     1.3  \nl
   &    29.0-48.3\arcsec\ SW &  2.4$\times$10$^{-17}$     &  114   &    2.1  &     1.8  \nl
   &    49.0-69.0\arcsec\ SW &  1.6$\times$10$^{-17}$   &    79   &    1.6  &     1.4  \nl
   &    69.1-100.7\arcsec\ SW &  1.5$\times$10$^{-17}$    &   64   &    1.1  &     1.1  \nl
   &    105.6-126.2\arcsec\ SW &   1.6$\times$10$^{-17}$   &    77   &    1.0 &      1.2  \nl
   &    131.8-141.4\arcsec\,+144.3-151.8\arcsec\ SW &  1.4$\times$10$^{-17}$   &    56   &   0.69  &    0.83  \nl

\nl

M 101 & 144.8-140.0\arcsec\ SE &  4.0$\times$10$^{-17}$  &     48   &   0.57  &    0.57  \nl
   &   135.8-120.0\arcsec\ SE  &  3.6$\times$10$^{-17}$    &   35  &    0.51  &    0.73  \nl
   &   111.7-86.9\arcsec\ SE &  2.4$\times$10$^{-17}$  &     43   &   0.61  &    0.72  \nl
   &   77.2-66.2\arcsec\ SE  &  3.1$\times$10$^{-17}$  &     40   &   0.74  &    0.66  \nl
   &   64.1-60.7\arcsec\ SE &  5.1$\times$10$^{-17}$   &     45  &    0.49  &    0.47  \nl
   &   53.7-46.2\arcsec\ SE &  4.2$\times$10$^{-17}$   &    44   &   0.69   &   0.54  \nl
   &   42.7-40.7\arcsec\ SE &  4.6$\times$10$^{-17}$  &     27  &    0.61   &   0.39  \nl
   &   29.6-20.7\arcsec\ SE &  3.7$\times$10$^{-17}$   &    79  &     1.2   &   0.79  \nl
    &   8.4-15.2\arcsec\ NW  & 2.9$\times$10$^{-17}$   &    77  &     2.3   &     1.3  \nl
   &    15.3-20.7\arcsec\ NW  & 1.9$\times$10$^{-17}$   &    89   &    2.7   &    1.2  \nl
   &    22.2-28.3\arcsec\ NW &  3.5$\times$10$^{-17}$   &    83   &   0.94   &   0.34  \nl
   &     35.3-45.6\arcsec\ NW &  4.0$\times$10$^{-17}$   &    36  &    0.55  &    0.36  \nl
   &     49.1-56.6\arcsec\ NW &  4.2$\times$10$^{-17}$    &   34   &   0.55  &    0.48  \nl
   &    80.2-84.2\arcsec\ NW &  4.3$\times$10$^{-17}$   &    47     & 0.45  &    0.45  \nl
   &    86.4-117.3\arcsec\ NW &  1.3$\times$10$^{-17}$    &   45    &   1.1   &    1.1  \nl
    &   117.4-144.2\arcsec\ NW &  2.1$\times$10$^{-17}$     &     40   &   0.64   &   0.69  \nl

\nl

NGC 2403 & 145.1-130.0\arcsec\ NE &  3.4$\times$10$^{-17}$   &   51  & 0.27  &   0.75  \nl
   &   130.0-119.7\arcsec\ NE &  4.1$\times$10$^{-17}$  &     54 &    0.21  &    0.67  \nl
  &    74.8-67.3\arcsec\ NE &  5.5$\times$10$^{-17}$   &     56   &   0.30  &    0.62  \nl
   &   67.2-59.7\arcsec\ NE & 4.4$\times$10$^{-17}$    &    54   &   0.37  &    0.71  \nl
   &     87.4-103.2\arcsec\ SW  &  2.4$\times$10$^{-17}$   &  60   &   0.36  &      1.1  \nl
   &    103.3-151.5\arcsec\ SW &  1.7$\times$10$^{-17}$   &    59   &   0.37  &    0.93  \nl

\nl

NGC 4395 & 146.0-90.9\arcsec\ NE &  9.3$\times$10$^{-18}$   &    42   &   \nodata  &  \nodata  \nl
  &    90.8-79.2 \arcsec\ NE&  3.1$\times$10$^{-17}$  &     29  &    \nodata   &   \nodata  \nl
  &     79.1-39.2\arcsec\ NE &  1.6$\times$10$^{-17}$  &     47  &   \nodata  &  \nodata  \nl                      
  &    34.9-24.0\arcsec\ NE & 2.9$\times$10$^{-17}$  &     57 &   \nodata  &  \nodata  \nl                       
  &    21.1-10.9\arcsec\ NE  & 3.3$\times$10$^{-17}$  &     78   &   \nodata  &  \nodata  \nl                     
  &       7.8-15.3\arcsec\ SW &  2.3$\times$10$^{-17}$  &     46  &   \nodata  &  \nodata  \nl                      
  &     29.9-48.5\arcsec\ SW  & 2.5$\times$10$^{-17}$   &    50    &   \nodata  &  \nodata  \nl                    
  &     65.1-94.7\arcsec\ SW &  1.2$\times$10$^{-17}$  &      60    &   \nodata  &  \nodata  \nl                    
  &     94.8-150.6\arcsec\ SW &   6.7$\times$10$^{-18}$  &     41     &   \nodata  &  \nodata  \nl                   
  &     NE sum\tablenotemark{a}  & 1.7$\times$10$^{-17}$    &    53   &   0.14  &    0.66 \nl
  &     SW sum\tablenotemark{a}  & 1.2$\times$10$^{-17}$    &    56   &   0.19  &    0.67 \nl

\tablecomments{1) Col. 2---aperture is listed as the offset from the nucleus
        in arcseconds (the slit is 2\arcsec\ wide); 
        Col.~3---The surface brightness 
        of H$\alpha$ emission line in erg s$^{-1}$ cm$^{-2}$ arcsecond$^{-2}$;
        Col. 4---FWHM in km s$^{-1}$ of H$\alpha$ line (in the case that
        EQW(H$\alpha$) $>$ 3\AA) and [NII]6583 (in the case that
        EQW(H$\alpha$) $<$ 3\AA)
        after correction for the instrumental resolution; 
        Col. 5 \& 6---[NII]$\lambda$6583/H$\alpha$ and 
        [SII]$\lambda$6716+$\lambda$6731/H$\alpha$ respectively. 
        2) The effect of the stellar H$\alpha$ 
        absorption has been corrected (see text for details). }

\tablenotetext{a}{`NE sum'---the listed apertures NE of the nucleus 
        were summed to 
        measure the [NII] and [SII] emission lines, due to the weakness of
        these lines in the spectrum;
        `SW sum'---similarly the listed apertures SW of the nucleus 
        were summed.}

\enddata

\end{planotable}



\begin{deluxetable}{lccccc}
\scriptsize
\tablewidth{0pt}
\tablecaption{Emission-line Properties of DIM Part II. [OIII] and [OI]}
\tablehead{
\colhead{Object }     &  \colhead{Aperture}   &  
\colhead{$\Sigma_{{\tt [OIII]5007}}$\tablenotemark{a}}  & \colhead{$\Delta V_{{\tt [OIII]5007}}$\tablenotemark{b}}  &
\colhead{[OIII]5007/H$\beta$\tablenotemark{c}}   &
\colhead{[OI]6300/H$\alpha$\tablenotemark{d}}  \\
\colhead{}  &  \colhead{}  & \colhead{(erg s$^{-1}$ cm$^{-2}$ arcsec$^{-2}$)} &
\colhead{(km s$^{-1}$)}  & \colhead{}  &  \colhead{}
}

\startdata

M 51    &  sum NE\tablenotemark{e} & 1.4$\times$10$^{-17}$  &  180  &  2 &  \nodata   \nl
    &   sum SW\tablenotemark{f}   & 7.6$\times$10$^{-18}$  &  130  &    1.5 & \nodata  \nl
     &  46.9-62.0\arcsec\ SW      & 5.9$\times$10$^{-18}$  &  100  &   1  &  \nodata   \nl
    &   \nodata\tablenotemark{d}  &  \nodata   &   \nodata   &  \nodata   & 0.1(marginal)    \nl
\nl

M 101  &  144.8-140.0\arcsec + 135.5-120.3\arcsec\ SE  &  $<$2.4$\times$10$^{-18}$  &  \nodata  &  $<$0.2  &   \nodata  \nl
     &   77.2-66.2\arcsec + 63.8-61.0\arcsec\ SE    &  7.2$\times$10$^{-18}$  &  130  &  0.3   &    \nodata   \nl
     &   53.7-46.2\arcsec + 42.4-41.0\arcsec\ SE    &  $<$4.7$\times$10$^{-18}$  &  \nodata  &  $<$0.4   &    \nodata  \nl
     &   22.2-28.3\arcsec\ NW                       &  1.3$\times$10$^{-17}$  &  130  &  0.6   &    \nodata  \nl
     &   35.3-45.6\arcsec + 49.4-56.3\arcsec\ NW    &  8.1$\times$10$^{-18}$  &  110  &  0.4    &    \nodata  \nl
     &   80.2-84.2\arcsec\ NW                         &  8.0$\times$10$^{-18}$  &  50  &  0.7   &    \nodata   \nl
     &  \nodata\tablenotemark{d}   &  \nodata   &  \nodata   & \nodata   &   0.1(marginal)   \nl
\nl

NGC 2403 & 145.1-130.0\arcsec\ NE &  1.6$\times$10$^{-17}$  &  100  & 1.5  &  \nodata  \nl
   &   130.0-119.7\arcsec\ NE &  1.7$\times$10$^{-17}$  &  50   & 1.4  &   \nodata  \nl
  &    74.8-67.3\arcsec\ NE &  1.5$\times$10$^{-17}$  &  170  & 1.0   &  \nodata  \nl
   &   67.2-59.7\arcsec\ NE &  1.3$\times$10$^{-17}$   &  160  & 1.0   &  \nodata  \nl
   &     87.4-151.5\arcsec\ SW  &  3.6$\times$10$^{-18}$  &  60  & 0.65   &  \nodata  \nl
  &   \nodata\tablenotemark{d}  &  \nodata   & \nodata   & \nodata   &   0.2    \nl

\nl

NGC 4395  &  34.9-24.0\arcsec\ NE  &  7.7$\times$10$^{-18}$  &  $<$40  & 0.7   & \nodata  \nl
  &    nearnuc sum \tablenotemark{g} &  1.5$\times$10$^{-17}$  &  170  & 2.4  &  \nodata  \nl
  &  29.9-48.5\arcsec\ SW &  $<$3.2$\times$10$^{-18}$   & \nodata  &  $<$0.4    &  \nodata   \nl

\tablebreak

\tablecomments{1) Most apertures listed in Col. 2 are defined in the 
        same way as in Table 3. Apertures were selectively summed 
        as necessary to achieve the best S/N ratio. Some apertures 
        associated with very noisy spectra were not used. 
        2) The effect of the stellar H$\alpha$ 
        absorption has been corrected (see text for details).}

\tablenotetext{a}{[OIII]$\lambda$5007 line surface brightness in units of erg s$^{-1}$ cm$^{-2}$ arcsecond$^{-2}$. In the regions where the line is not detected, upper limits are given which are estimated using 5 $\sigma$ of
 background.}
\tablenotetext{b}{Deconvolved line width of [OIII]$\lambda$5007
in km/s. In the regions where the line is not resolved, 
upper limits are given which are estimated using the instrument resolution.}
\tablenotetext{c}{[OIII]$\lambda$5007/H$\beta$ ratios. 
In the regions where the [OIII]$\lambda$5007 line is not detected, 
upper limits are given which are estimated using 5 $\sigma$ of
 background for [OIII]$\lambda$5007.}
\tablenotetext{d}{The apertures used to measure [OI]$\lambda$6300 line for 
        M51 and M101 are
        the sum of the individual apertures listed in Table 3.
        For NGC2403 the apertures 134.1-119.7\arcsec\ NE and
        87.4-122.5\arcsec\ SW were summed. The [OI]$\lambda$6300 line is 
        blended with night sky lines in M81 and NGC4395 spectra and was 
        thus not measurable.}
\tablenotetext{e}{The apertures summed are 92.5-90.5\arcsec\ NE, 
        84.9-80.8\arcsec\ NE and 63.5-52.6\arcsec\ NE.}
\tablenotetext{f}{The apertures summed are 95.2-102.0\arcsec\ SW,
        105.5-107.5\arcsec\ SW and 117.3-152.4\arcsec\ SW. }
\tablenotetext{g}{The aperture summed are 21.1-10.9\arcsec\ NE and
        7.8-15.3\arcsec\ SW. Note that NGC 4395 has a Seyfert nucleus, 
        which may affect the [OIII]/H$\beta$ ratio and [OIII]5007 line width.}

\enddata

\end{deluxetable}



\begin{deluxetable}{lcrccc}
\tablewidth{34pc}
\tablecaption{DIM Imaging Result\tablenotemark{a}}
\tablehead{
\colhead{Object}   &
\colhead{H$\alpha$ Flux}          & \colhead{L$_{{\tt H}\alpha}$ } &
\colhead{Mean $\Sigma_{{\tt H}\alpha}$\tablenotemark{b}} & 
\colhead{Mean/r.m.s.\tablenotemark{c}}    &
\colhead{$\Sigma_{50}$\tablenotemark{d}} \\
\colhead{} & \colhead{(erg s$^{-1}$ cm$^{-2}$)} & \colhead{(L$\sun$)} & 
\colhead{(pc cm$^{-6}$)} & \colhead{} & \colhead{(pc cm$^{-6}$)}
}
\startdata
M 51  &  2.0$\times$10$^{-11}$  &  4.4$\times$10$^{7}$  & 35  &  0.24  &  400    \nl
M 81  &  3.2$\times$10$^{-11}$  &  1.3$\times$10$^{7}$  & 16  &  0.30  &  74    \nl
M 101 &  3.7$\times$10$^{-11}$  &  6.3$\times$10$^{7}$  & 9.5  &  0.12  & 230   \nl
NGC 2403 & 3.3$\times$10$^{-11}$  &  1.1$\times$10$^{7}$  &  19  &  0.19  &  230  \nl
NGC 4395 & 9.2$\times$10$^{-12}$ & 1.9$\times$10$^{6}$  & 9.2 & 0.24  &  74   \nl

\tablenotetext{a}{Correction for foreground Galactic extinction has not been
  applied to the data in this table. }
\tablenotetext{b}{Mean H$\alpha$ surface brightness within the galaxy's
  B = 25 mag arcsec$^{-2}$ isophote converted to emission
  measure assuming an electron temperature of 10$^4$ K
  (5$\times$10$^{-17}$ erg s$^{-1}$ cm$^{-2}$ arcsec$^{-2}$ 
  corresponds to 25 pc cm$^{-6}$). }
\tablenotetext{c}{The ratio of mean to r.m.s of H$\alpha$ surface brightness.
  This ratio is used to characterize the `DIMness' of a galaxy. See text for
  more details. }
\tablenotetext{d}{H$\alpha$ surface brightness level at which fainter gas
  contributes 50\% of total H$\alpha$ flux. The unit is converted to 
  that of emission measure assuming an electron temperature of 10$^4$ K. }

\enddata
\end{deluxetable}



\begin{planotable}{lccccc}
\tablewidth{38pc}
\tablecaption{Observed Typical Emission Line Ratios of the DIM}
\tablehead{
\colhead{Object}      & \colhead{[NII]$\lambda$6583/H$\alpha$} &
\colhead{[SII]$\lambda$6717/H$\alpha$\tablenotemark{a}} & \colhead{[OI]$\lambda$6300/H$\alpha$\tablenotemark{b}}    &
\colhead{[OIII]$\lambda$5007/H$\beta$} }

\startdata

M 51    &   1.2; 0.44    &   0.47; 0.11   &  0.1($^*$); 0.015   &   1; 0.4  \nl
M 81    &   1.6; 0.53    &   1.1; 0.43  &  ...   &   ...  \nl
M 101    &   0.64; 0.26    &   0.36; 0.13   &  0.1($^*$); 0.012   &   0.4; 0.1  \nl
NGC 2403    &   0.29; 0.28    &   0.45; 0.21   &  0.2; 0.030   &   1.0; 0.4  \nl
NGC 4395    &   0.19; 0.11    &   0.42; 0.14   &  ...   &   0.6; 2  \nl
Galaxy\tablenotemark{c}    &    0.3;\nodata    &     0.35;\nodata    &     $<$0.02;\nodata    &     $<$0.2;\nodata   \nl
NGC 891 reg. 1\tablenotemark{d}   &   0.6-1.1;\nodata     &   0.3-0.6;\nodata     &  $<$0.05;\nodata     &   $<$0.4;\nodata \nl
NGC 891 reg. 2\tablenotemark{d}   &   0.4-0.8;\nodata   &  0.25-0.35;\nodata     &    \nodata    &   \nodata    \nl

\tablecomments{Listed line-ratios are from averaged DIM and 
        HII region spectra. Semi-colon separated values are for the DIM and
        HII regions respectively; For the Galaxy and NGC 891 only the DIM
        line-ratios are given. }
\tablenotetext{a}{ [SII]$\lambda$6731 is not included in order to compare
         to the data of the Milky Way and NGC 891. Including this line would
         increase the ratios by a factor $\simeq$1.7. }
\tablenotetext{b}{ Ratios marked with ($^*$) are calculated from the marginally
  detected [OI]$\lambda$6300 line flux. }
\tablenotetext{c}{ From Reynolds 1990, 1991. }
\tablenotetext{d}{ From Dettmar 1992. }

\enddata

\end{planotable}



\begin{deluxetable}{lccccc}
\tablewidth{0pt}
\tablecaption{DIM Scale-Heights}
\tablehead{
\colhead{Object}   &
\colhead{$\Sigma_*$\tablenotemark{a}}  & \colhead{$\Delta V_{{\tt H}\alpha,{\tt NII}}$\tablenotemark{b} } &
\colhead{H$_{{\tt H}\alpha,{\tt NII}}$\tablenotemark{c}} & 
\colhead{$\Delta V_{{\tt OIII5007}}$\tablenotemark{d}} &
\colhead{H$_{{\tt OIII5007}}$\tablenotemark{e}} \\
\colhead{} & \colhead{(M$_{\sun}$\ pc$^{-2}$)} & \colhead{(km s$^{-1}$)} & 
\colhead{(pc)} & \colhead{(km s$^{-1}$)} & \colhead{(pc)}
}
\startdata

M 51  &  200  &  80-100 & 470-620 & 110-170  &  700-1300  \nl
M 81   &  150  &  60-150 & 400-1300    & \nodata  &  \nodata \nl
M 101  &  340  &  30-90    & 120-400   &  60-120 &  250-560   \nl
NGC 2403 & 250 &  50-60   & 240-300    &  60-160 &  300-1000      \nl
NGC 4395   &  94  &  40-60 &  330-530   & \nodata  &  \nodata \nl

\tablenotetext{a}{Surface density of stellar mass within the central 5\arcmin\ 
  regions of the galaxies that are covered by our spectroscopic observations.
  These values are estimated from V- and B-band surface photometry 
  measurements by 
  Okamura et al. (1976)(M 51 and M 101), Brandt et al. (1972) (M 81), 
  Bothun \& Rogers (1992) 
  (NGC 2403) and van der Kruit (1987) (NGC 4395). 
  A correction for inclination has been applied and a mass-to-light
  ratio of 5 times the solar value in both bands has been adopted
  in this calculation. }

\tablenotetext{b}{The range of the FWHMs of the DIM emission lines 
  H$\alpha$ (for cases in which the H$\alpha$ equivalent width EQW(H$\alpha$)
  is larger 
  than 3$\AA$) and [NII]6583 
  (when the EQW(H$\alpha$) is less than 3$\AA$).
  The observed values are equated to those in the $z$ direction by assuming
  isotropic motion of the gas. }
\tablenotetext{c}{Scale height of the diffuse gas corresponding to the FWHMs
  of the H$\alpha$ and [NII] lines, calculated assuming an
  exponential distribution of mass in the $z$ direction with a scale height
  of 350 pc. See text for further details. }

\tablebreak

\tablenotetext{d}{The range of the FWHM of the DIM emission line
  [OIII]$\lambda$5007.
  The observed values are equated to those in the $z$ direction by assuming
  isotropic motion of the gas. }
\tablenotetext{e}{Scale height of the diffuse gas corresponding to the FWHMs
  of the [OIII]$\lambda$5007 line, calculated assuming an
  exponential distribution of mass in the $z$ direction with a scale height
  of 350 pc. See text for further details. }

\enddata
\end{deluxetable}


\end{document}